\documentclass[journal]{IEEEtran}
\usepackage{amsmath}
\usepackage{amsmath,bm}
\usepackage{amsfonts} 
\usepackage{graphicx} 
\usepackage{subcaption}
\usepackage{array}

\usepackage{enumitem}
\usepackage{multirow}

\usepackage{diagbox}

\usepackage[T1]{fontenc}
\usepackage{authblk}

\begin{document}
\title{A Novel Convolutional Neural Network for Image Steganalysis with Shared Normalization}

%

\author[1,2]{Songtao Wu}
\author[*,1]{Sheng-hua Zhong}
\author[2]{Yan Liu}
\affil[1]{College of Computer Science and Software Engineering, Shenzhen University }
\affil[2]{Department of Computing, The Hong Kong Polytechnic University}
\affil[ ]{\textit{csstwu@szu.edu.cn, csshzhong@szu.edu.cn, csyliu@comp.polyu.edu.hk}}

\markboth{SUBMISSION TO IEEE TRANSACTION ON MULTIMEDIA}%
{Shell \MakeLowercase{\textit{et al.}}: Bare Demo of IEEEtran.cls for IEEE Journals}
\maketitle

\begin{abstract}
   Deep learning based image steganalysis has attracted increasing attentions in recent years. Several Convolutional Neural Network (CNN) models have been proposed and achieved state-of-the-art performances on detecting steganography. In this paper, we explore an important technique in deep learning, the batch normalization, for the task of image steganalysis. Different from natural image classification, steganalysis is to discriminate cover images and stego images which are the result of adding weak stego signals into covers. This characteristic makes a cover image is more statistically similar to its stego than other cover images, requiring steganalytic methods to use paired learning to extract effective features for image steganalysis. Our theoretical analysis shows that a CNN model with multiple normalization layers is hard to be generalized to new data in the test set when it is well trained with paired learning. To hand this difficulty, we propose a novel normalization technique called Shared Normalization (SN) in this paper. Unlike the batch normalization layer utilizing the mini-batch mean and standard deviation to normalize each input batch, SN shares same statistics for all training and test batches. Based on the proposed SN layer, we further propose a novel neural network model for image steganalysis. Extensive experiments demonstrate that the proposed network with SN layers is stable and can detect the state of the art steganography with better performances than previous methods.
\end{abstract}

\IEEEpeerreviewmaketitle

\section{Introduction}
\IEEEPARstart{S}{teganography} is the technique to hide secret messages into multimedia signals such as audio, image or video [1]. Steganalysis, from an opponent's perspective, is the art of revealing the presence of secret messages embedded in the digital medias [2]. Among all steganalytic techniques, image steganalysis plays an important role in many information security systems and attracts increasing interests in recent years [3-4].

Designing effective features that are sensitive to message embedding is key to image steganalysis. Traditional feature-based methods, Spatial Rich Model (SRM) [5] steganalysis and its selection-channel version [6], assemble various handcrafted features and learn a binary classifier to detect steganography. However, designing effective features turns out to be a difficult task which needs strong domain knowledge of steganography and steganalysis. Recently, several interesting works have been proposed to detect steganography based on deep convolutional neural networks [7-10, 14-17]. Compared with traditional methods that extract handcrafted features, CNN based steganalysis automatically learns effective features using various network architectures for discriminating cover images and stego images. Tan and Li [7] first proposed to detect the presence of secret messages based on a deep stacked convolutional auto-encoder network. Qian \textit{et al.} [8] proposed a model for steganalysis using the standard CNN architecture with Gaussian activation function and further proved that transfer learning is beneficial for a CNN model to detect a steganographic algorithm at low payloads in [9]. Xu \textit{et al.} [10] designed a compact and effective CNN architecture with multiple batch normalization layers. This network now becomes a basic model for several more complex CNN models [11-13]. Wu \textit{et al.} [14-15] proposed a novel CNN model to detect steganography based on residual learning and achieved low detection error rates when cover images and stego images are paired. Ye \textit{et al.} in [16] firstly incorporated the truncation technique into the design of steganalytic CNN models. With the help of selection channel knowledge and data augmentation, their model obtained significant performance improvements than the classic SRM on resampled and cropped images. Except for the spatial domain steganalysis, Xu \textit{et al.} [17] proposed a network based on residual learning and achieved better detection accuracy than traditional methods in compressed domain.


Training deep neural network is difficult since data at different layers follow different distributions [18]. Batch normalization is a standard technique widely used in many CNN models [19-20], which can effectively handle this difficulty. By normalizing input data with mini-batch statistics, elements in feature maps are forced to be distributed with zero mean and unit standard deviation. Batch normalization not only reduces the training time of CNN models significantly, but also regularizes the network implicitly. As an important component of a CNN model, batch normalization has shown promising performances in different image related tasks, such as classification [19], image denoising [20], and image captioning [21].

Steganalysis is different from natural image classification. It is challenging since steganalysis aims to discriminate cover images and stego images which are the results of adding very weak stego signals into covers. This characteristic leads to the result that a cover image is more statistically similar to its stego image than other cover images. In order to learn discriminative features between cover images and stego images, existing steganalytic methods [5-6, 8-17] often use paired training, i.e. cover images and their stegos should be both in the training set, for steganalytic feature extraction. In this paper, we analyze that how batch normalization affects the performance of CNN-based image steganalysis. Our theoretical analysis indicates that the batch normalization can capture the characteristic of paired learning, which makes the model fall into a dilemma. For one hand, a CNN model with batch normalization layers can detect steganography accurately once training batch and test batch have consistent statistical properties, i.e. cover images and their stegos are paired both in training set and test set. However, the requirement of paired learning and paired testing is too restrict to be satisfied in real applications. For the other hand, the model trained with paired learning is hard to be generalized to the test data once cover images and their stegos are not paired in a test batch.

To address the limitation of batch normalization for image steganalysis, a novel normalization technique called Shared Normalization (SN) is proposed. Unlike the batch normalization that use mini-batch mean and standard deviation to normalize the data, SN shares same statistics for all training and test batches. With unified statistics, CNN models are forced to learn features to discriminate cover images and stego images, thus improves the model's generalization ability. Based on the proposed SN technique, we further introduce a novel CNN model to detect steganography. Our CNN model recursively uses a processing unit, which contains a convolutional layer, a SN layer, a ReLU layer and an average pooling layer, to extract effective features for image steganalysis. Experiments demonstrate that the proposed CNN model can not only learn more effective features compared with classical CNN architecture, but also detect steganoraphic algorithms with better performance than SRM and other CNN models.


The rest of this paper is organized as follows. In section II, we introduce the preliminary knowledge about convolutional neural network and batch normalization. In section III, we analyze the limitation of batch normalization for image steganalysis in theory. In section IV,  we propose a novel normalization technique called shared normalization and further introduce a novel convolutional neural network model based on the proposed SN for image steganalysis. In section V, we validate the effectiveness of the proposed model on several states of the art steganographic algorithms. The paper is finally closed with the conclusion in section VI.

\section{Preliminaries}
\subsection{Rationality of Using Convolutional Neural Network for Image Steganalysis}
In recent years, CNN has achieved great success in many image related tasks. A series of breakthroughs have been made for discriminative learning, including image classification [19, 22-23], image denoising [20], and image super-resolution [24]. In addition, several recent work show that CNN models are successfully applied for generative learning, including real image generation [25] and texture synthesis [26]. These successes indicate that CNN can not only extract effective features for discriminating different images but also provide a good description for representing real images. All the evidences show that CNN can well describe the distribution of natural images. These results motivate us to use CNN for image steganalysis, since its purpose is to discriminate the ``natural" images (cover images) against the ``unnatural" images contaminated by embedded secret messages (stego images). The following three characteristics of CNN models further demonstrate that they are suitable for the task of image steganalysis:
\begin{itemize}
  \item Convolutional kernels in CNN models can exploit the strong spatially local correlation present in input images [27]. This local correlation among image pixels is distorted when secret messages are embedded, making it different from the correlation in natural images. The difference between natural images and distorted images can be effectively captured by CNN models;
  \item The convolution operation is actually to sum image pixels in a local region, which would accumulate the weak stego signal of this region to be a large value. This may lead to stego images be more easily detected against cover images;
  \item Nonlinear mappings in CNN models make them able to extract rich features for classifying cover images and stego images. These features, which are automatically learned by updating the network, can hardly be designed by hand.
\end{itemize}
\subsection{Batch Normalization for Convolutional Neural Network}
Batch normalization is a standard technique that is widely used in CNN models for image classification [18]. Training a deep neural network model is often difficult not only because of the gradient vanishing/exploding but also because the distribution of data changes in different layers, which is called the ``internal covariate shift" phenomenon. Batch normalization is such a technique that can relieve this phenomenon, by introducing several simple operations to the input data:
\begin{equation}
  \mu_{\mathcal{B}} \leftarrow \frac{1}{m}\sum_{i=1}^{m} I_{i}
\end{equation}
\begin{equation}
  \sigma_{\mathcal{B}}^{2} \leftarrow \frac{1}{m}\sum_{i=1}^{m} \left( I_{i} -  \mu_{\mathcal{B}} \right)^{2}
\end{equation}
\begin{equation}
  \hat{I}_{i} = \frac{I_{i}-E\left[I_{i}\right]}{\sqrt{\sigma_{\mathcal{B}}^{2} + \epsilon}}
\end{equation}
\begin{equation}
  I_{i}^{o} = \gamma \hat{I}_{i} + \beta
\end{equation}
where $I_{i}$ denotes the $i$-th training sample, $m$ is the number of samples in the batch, $\mathcal{B} = \{ I_{1,...m} \}$ denotes the input data in a mini-batch, $\mu_{\mathcal{B}} $ and $\sigma_{\mathcal{B}}$ represents the mean and standard deviation of the mini-batch $\mathcal{B}$ respectively. $\epsilon$ is a small constant to avoid zero dividing, $\gamma$ and $\beta$ are the parameters. With these operations, the output data $I_{i}^{o}$ in the mini-batch is distributed with fixed mean and standard deviation at any depth after the batch normalization. Thus, deviations to the mean and variance can be eliminated by the batch normalization, which makes the network overcome the ``internal covariate shift".

\section{Limitation of Batch Normalization for Image Steganalysis}
In this section, we introduce the limitation of batch normalization for image steganalysis. Firstly, we introduce an important characteristic of blind steganalysis: paired training. Then, based on this characteristic, we analyze that a CNN model with multiple batch normalization layers is hard to be generalized when it is trained by paired learning.

\subsection{Paired Training in Image Steganalysis}
Steganalysis is usually formulated as a binary classification problem [8-9]. This technique, which is called ``universal/blind steganalysis", becomes the main stream among most current steganalytic algorithms. In the training phase, effective features sensitive to message embedding are extracted to highlight potential manipulation by steganographier. Then, a binary classifier is learned on pairs of cover images and their corresponding stegos aiming to find a boundary to detect steganography. In testing phase, the trained classifier is used to predict labels of new input images. Previous research [29] showed that it is rather important to force cover features and stego features to be paired, i.e. steganalytic features of cover images and their stego images should be preserved in the training set. Otherwise, breaking cover-stego pairs in different sets would introduce biased error and lead to a suboptimal performance [30]. Mainstream steganalytic methods use paired training to learn effective features for image steganalysis [8].

\subsection{Paired Training and Paired Testing for a CNN Model with Multiple Batch Normalization Layers}
Generally, in a CNN model with multiple batch normalization layers, a batch normalization layer (BN) is after a convolutional layer (Conv) and before a ReLU activation layer. This ``Conv+BN+ReLU" block is a standard setting in many convolutional neural networks [19, 31-32]. For this configuration, we find that cover images and their stegos can be easily classified when they are fed into the block in pair and the batch statistics are used to normalize the data. To illustrate this phenomenon, we provide mathematical analysis here. Assume the network is fed with a batch which only contains a cover image $\mathbf{x}$ and its stego image $\mathbf{y}$:
\begin{equation}
 \mathbf{y} = \mathbf{x} + \mathbf{s}
\end{equation}
where $\mathbf{s}$ denotes stego signal introduced by message embedding. For the block ``Conv+BN+ReLU", the output of the cover image is:
\begin{small}
\begin{equation}
   \mathbf{x}^{op} = \text{ReLU}\left( \frac{\mathbf{W}\mathbf{x}-\mu}{\sigma}\right)  = \left[\frac{\mathbf{W}\mathbf{x}-\mu}{\sigma}\right] \circ \mathcal{H}\left[\frac{ \mathbf{Wx} - \mu}{\sigma} \right]
\end{equation}
\end{small}where $\mathbf{x}^{op}$ represents the output of cover image in paired case respectively. $\mu$ denotes the mean value of all pixels in $\mathbf{x}$ and $\mathbf{y}$, $\sigma$ represents its standard deviation. $\circ$ represents the pointwise product, $\mathcal{H}(\cdot)$ is Heaviside step function:
\begin{eqnarray}
  \mathcal{H}(x) =
   \left\{
   \begin{array}{lll}
     1, \ \  x \geq 0 \\
     0, \ \  x < 0
   \end{array}
   \right.
\end{eqnarray}
For simplicity, we have omitted the bias term and the scaling term in the batch normalization layer. Similarly, the output of stego image $\mathbf{y}^{op}$ is:
\begin{equation}
 \begin{aligned}
   \mathbf{y}^{op} & = \text{ReLU}\left( \frac{\mathbf{W}\mathbf{y}-\mu}{\sigma}\right) \\ & = \left[\frac{\mathbf{W}(\mathbf{x+s})-\mu}{\sigma}\right] \circ \mathcal{H}\left[\frac{ \mathbf{W}(\mathbf{x+s}) - \mu}{\sigma} \right]
 \end{aligned}
\end{equation}
For $\mu$, it is the mean of the cover image and stego image:
\begin{equation}
   \mu = \frac{1}{2}E\left[ \mathbf{W}\mathbf{x}+\mathbf{W}\mathbf{y} \right] = E\left[ \mathbf{W}\mathbf{x} \right] + \frac{1}{2}E\left[ \mathbf{W}\mathbf{s} \right]
\end{equation}
where $E[\cdot]$ denotes the expectation operator. For Eq.(6) and Eq.(8), we consider the expectation of batch normalization layer's outputs:
\begin{small}
\begin{equation}
   E\left[ \frac{\mathbf{W}\mathbf{x}-\mu}{\sigma} \right] =  E\left[ \frac{\mathbf{W}\mathbf{x}-E\left[ \mathbf{W}(\mathbf{x} + \frac{1}{2}\mathbf{s}) \right]}{\sigma} \right] = -\frac{E\left[\mathbf{W}\mathbf{s}\right]}{2\sigma}
\end{equation}
\end{small}
\begin{small}
\begin{equation}
   E\left[ \frac{\mathbf{W}\mathbf{y}-\mu}{\sigma} \right] =  E\left[ \frac{\mathbf{W}\mathbf{y}-E\left[ \mathbf{W}(\mathbf{x} + \frac{1}{2}\mathbf{s}) \right]}{\sigma} \right] = \frac{E\left[\mathbf{W}\mathbf{s}\right]}{2\sigma}
\end{equation}
\end{small}Eq.(10) and Eq.(11) indicate that the feature map of $\mathbf{x}$ and the feature map of $\mathbf{y}$  are distributed across 0 on average. This property makes elements in the cover feature map and elements in the stego feature map be automatically separated after the ReLU layer, once the $\mathbf{W}$ has been optimized to discriminate cover images and their stegos. Consequently, the classification of cover images and stego images becomes easy for a CNN model with ``Conv+BN+ReLU" block.

\begin{figure}[t]
   \centering
   \begin{subfigure}{.45\textwidth}
      \centering
      \includegraphics[height=6.4cm, width=8cm]{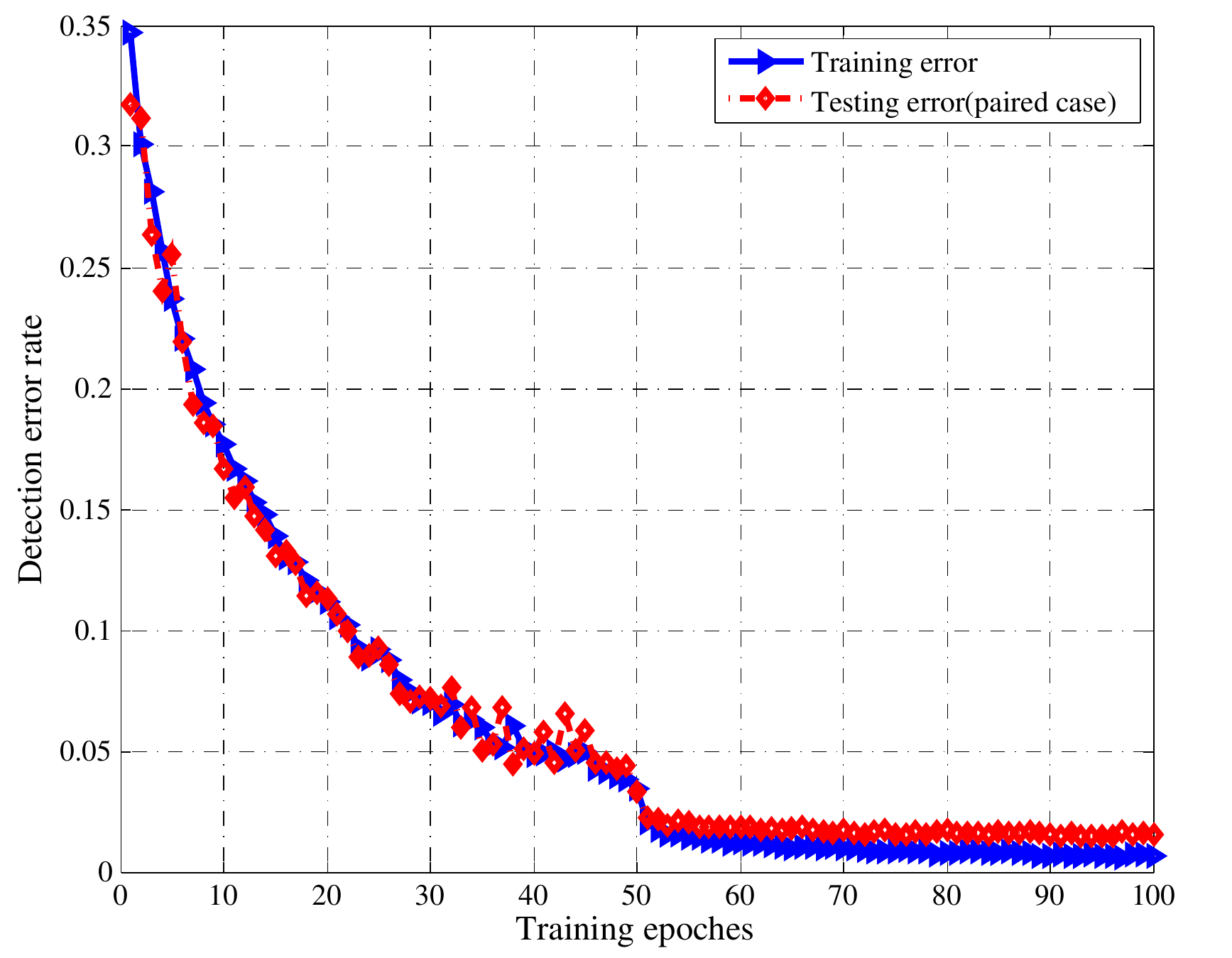}
   \end{subfigure}
      \caption{ Training error and detection error when cover images and stego images are paired in training and testing. The residual network proposed in [15] and S-UNIWARD at 0.4 bpp  are used for demonstration. The batch statistics are used to normalize input data both in training and testing. }
\end{figure}

For a network with many ``Conv+BN+ReLU" blocks, cover images and stego images are also easily classified when they are paired in a batch. In the paired learning case, the output of a cover image and its stego image is:
\begin{equation}
\mathbf{x}^{op}_{n} = \mathbf{f}^{pc}_{n} \circ \mathcal{H}\left[ \mathbf{f}^{pc}_{n}\right]
\end{equation}
\begin{equation}
\mathbf{y}^{op}_{n}  =  \mathbf{f}^{ps}_{n}  \circ \mathcal{H}\left[ \mathbf{f}^{ps}_{n} \right]
\end{equation}
where $\mathbf{x}^{op}_{n}$ and $\mathbf{y}^{op}_{n}$ represents the output of cover image and stego image after $n$-th ``Conv+BN+ReLU" blocks respectively. $\mathbf{f}^{pc}_{n}$ and $\mathbf{f}^{ps}_{n}$ represents the feature map of cover image and stego image after the batch normalization layer:
\begin{equation}
\mathbf{f}^{pc}_{n}= \frac{1}{\sigma_{n}} \left[ \mathbf{W}_{n}\mathbf{x}^{op}_{n-1} - \frac{1}{2}E\left( \mathbf{W}_{n}\mathbf{x}^{op}_{n-1} +\mathbf{W}_{n}\mathbf{y}^{op}_{n-1}  \right) \right]
\end{equation}
\begin{equation}
\mathbf{f}^{ps}_{n}= \frac{1}{\sigma_{n}} \left[  \mathbf{W}_{n}\mathbf{y}^{op}_{n-1} - \frac{1}{2}E\left( \mathbf{W}_{n}\mathbf{x}^{op}_{n-1} +\mathbf{W}_{n}\mathbf{y}^{op}_{n-1} \right) \right]
\end{equation}
where $\mathbf{W}_{n}$ denotes the $n$-th convolution kernel and $\sigma_{n}$ is the standard deviation. By introducing a variable defined as:
\begin{equation}
   \mathbf{s}_{n-1}^{op} = \mathbf{y}^{op}_{n-1} - \mathbf{x}^{op}_{n-1}
\end{equation}
Obviously, $\mathbf{s}_{n-1}^{op}$ is a weak noise introduced by the stego signal $\mathbf{s}$. For Eq.(14) and Eq.(15), we take expectation to $\mathbf{f}^{pc}_{n}$ and $\mathbf{f}^{ps}_{n}$ and obtain:
\begin{equation}
   E\left[ \mathbf{f}^{pc}_{n} \right] = -\frac{E\left[ \mathbf{W}_{n}\mathbf{s}_{n-1}^{op} \right]}{2\sigma_{n}}
\end{equation}
\begin{equation}
   E\left[ \mathbf{f}^{ps}_{n} \right] = \frac{E\left[ \mathbf{W}_{n}\mathbf{s}_{n-1}^{op} \right]}{2\sigma_{n}}
\end{equation}
Similar to Eq.(10) and Eq.(11), elements in the feature map of the cover image and the stego image are distributed across 0, which can be easily classified after the ReLU layer.

To demonstrate the fact that a CNN model with many batch normalization layer can detect steganography easily when cover images and their stego are paired in a batch, we use the residual network proposed in [15] for validation. For the network, we use the batch statistics, i.e. batch mean and batch variance, to normalize feature maps both in training and testing. The Spatial UNIversal WAvelet Relative Distortion (S-UNIWARD) steganography [33] at 0.4 bit-per-pixel (bpp) is used for evaluation. All settings of the network can refer to [15]. Fig.1 shows the detection errors when cover images and their stego images are paired both in training and testing.

In summary, we analyze that a CNN model with multiple batch normalization layers can detect steganography at high accuracy when covers and their stegos are paired in training and testing. However, cover images and their stegos should be paired in testing is a restrict assumption that might not be satisfied in real applications. Hence, in the next subsection, we will analyze the behavior of the CNN model with batch normalization layers when testing samples are not paired.

\begin{figure}[t]
   \centering
     \includegraphics[height=6.4cm, width=8cm]{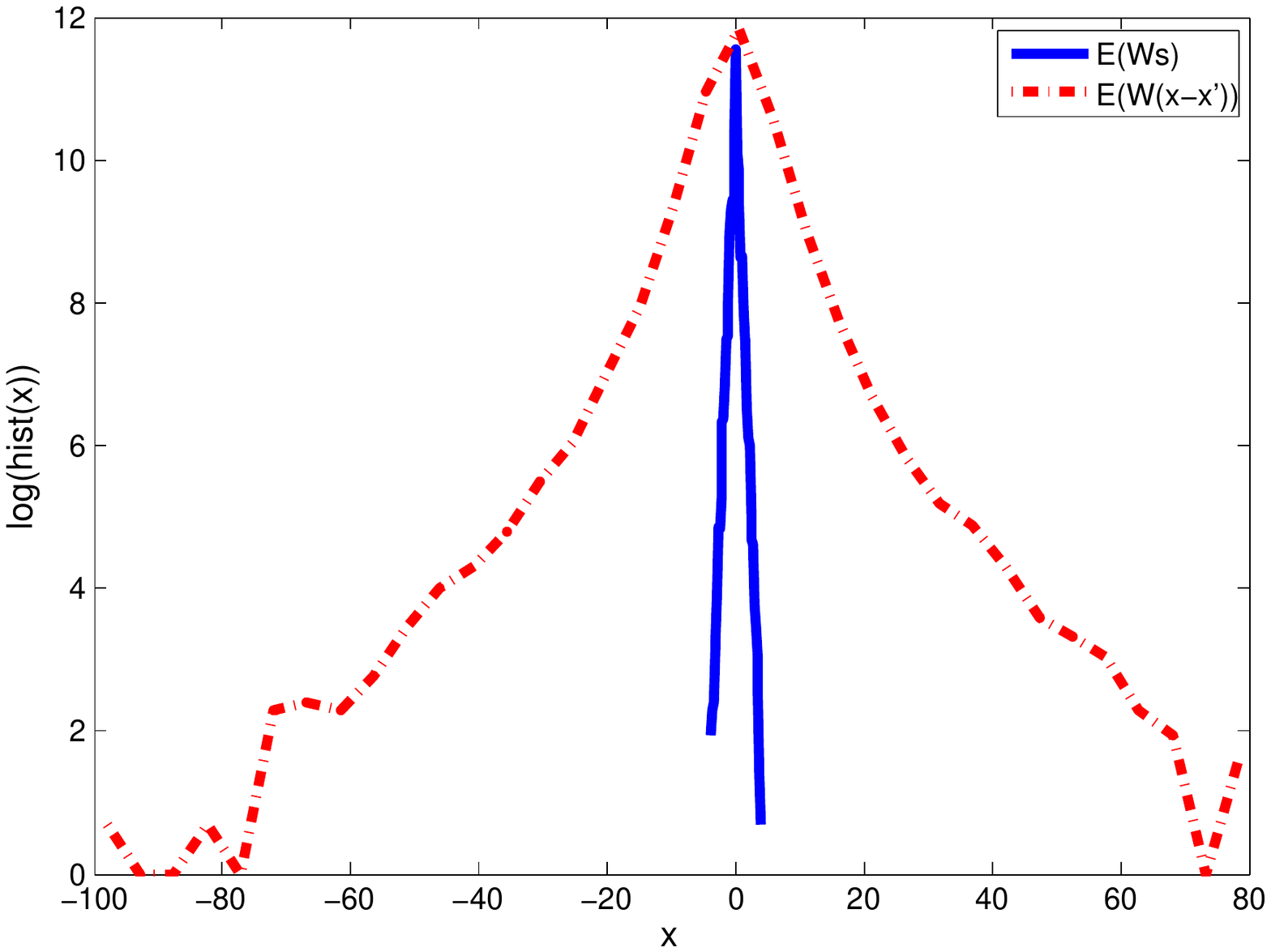}
     \caption{Histogram of elements in $\mathbf{Ws}$ and $\mathbf{W(x-x')}$. The residual network in [15] is used for validation. Elements in the feature map after high pass filtering are extracted for demonstration and the steganographic algorithm S-UNIWARD at payload 0.4 bpp is used for message hiding. }
\end{figure}

\subsection{Paired Training and Unpaired Testing for a CNN Model with Multiple Batch Normalization Layers}
In this subsection, we analyze the case that the network with batch normalization layers is trained by paired learning but tested with unpaired cover/stego samples. The following analysis demonstrates that the network can hardly predict the label of unpaired images if the batch statistics is used to normalize the data. Assume the block is fed with a cover image $\mathbf{x}$ and a stego image $\mathbf{y}'$, where $\mathbf{y}' = \mathbf{x}' + \mathbf{s}$ and $\mathbf{x}' \neq \mathbf{x}$. For the ``Conv+BN+ReLU" block, the outputs of $\mathbf{x}$ and $\mathbf{y}'$ are:
\begin{equation}
\mathbf{x}^{ou} = \frac{\mathbf{Wx} - \mu' }{\sigma'} \circ \mathcal{H}\left[\frac{ \mathbf{Wx} - \mu'} {\sigma'} \right]
\end{equation}
\begin{equation}
\mathbf{y}^{ou} = \frac{ \mathbf{W(x'+s)} - \mu' }{\sigma'} \circ \mathcal{H}\left[\frac{ \mathbf{W(x'+s)} - \mu'} {\sigma'} \right]
\end{equation}
where $\mathbf{x}^{ou}$ and $\mathbf{y}^{ou}$ denotes the output of cover image and stego image in unpaired case respectively, $\mu'$ and $\sigma'$ represents the mean and the standard deviation. $\mu'$ is:
\begin{equation}
  \mu' = \frac{1}{2}E\left( \mathbf{W}(\mathbf{x+x'+s}) \right)
\end{equation}
Similar to Eq.(10) and Eq.(11), we take expectation to the output of batch normalization layer in the unpaired case and obtain:
\begin{equation}
   E\left[ \frac{\mathbf{Wx} - \mu' }{\sigma'} \right] = \frac{1}{2\sigma'} \left[ E\left(\mathbf{W(x-x')}\right) - \frac{1}{2} E\left(\mathbf{Ws}\right) \right]
\end{equation}
\begin{equation}
   E\left[ \frac{\mathbf{Wy'} - \mu'}{\sigma'} \right] = \frac{1}{2\sigma'} \left[ E\left(\mathbf{W(x'-x)}\right) + \frac{1}{2} E\left(\mathbf{Ws}\right) \right]
\end{equation}
Compared with the paired learning case, on average, the expected output of batch normalization layer in Eq.(22) and Eq.(23) not only depends on $E\left[\mathbf{Ws}\right]$ but also on $E\left[\mathbf{W(x-x')}\right]$. Fig.2 has shown the distribution of elements in $\mathbf{W(x-x')}$ and $\mathbf{Ws}$. It is observed that the amplitude of secret message $\mathbf{Ws}$  is smaller than $\mathbf{W(x-x')}$. Therefore, in unpaired case, the output of the ``Conv+BN+ReLU" block is dominated by the difference $\mathbf{(x-x')}$ rather than the secret message $\mathbf{s}$. The result implies that the amplitude of the ``Conv+BN+ReLU" output is determined by cover images. For the model with multiple ``Conv+BN+ReLU" blocks, the iterative equations for $\mathbf{x}^{ou}$ and $\mathbf{y'}^{ou}$ at a deep depth are similar with Eq.(12)-Eq.(15) except that their difference is dominated by the term introduced by $\mathbf{(x-x')}$ at any depth. This interference term, however, cannot be eliminated as the network depth increases. Consequently, the feature map generated by the unpaired case completely different from the feature map generated by the paired case, thus the detection error rate would increase.
\begin{figure}[t]
   \centering
   \begin{subfigure}{.24\textwidth}
     \centering
     \includegraphics[height=3.3cm, width=4cm]{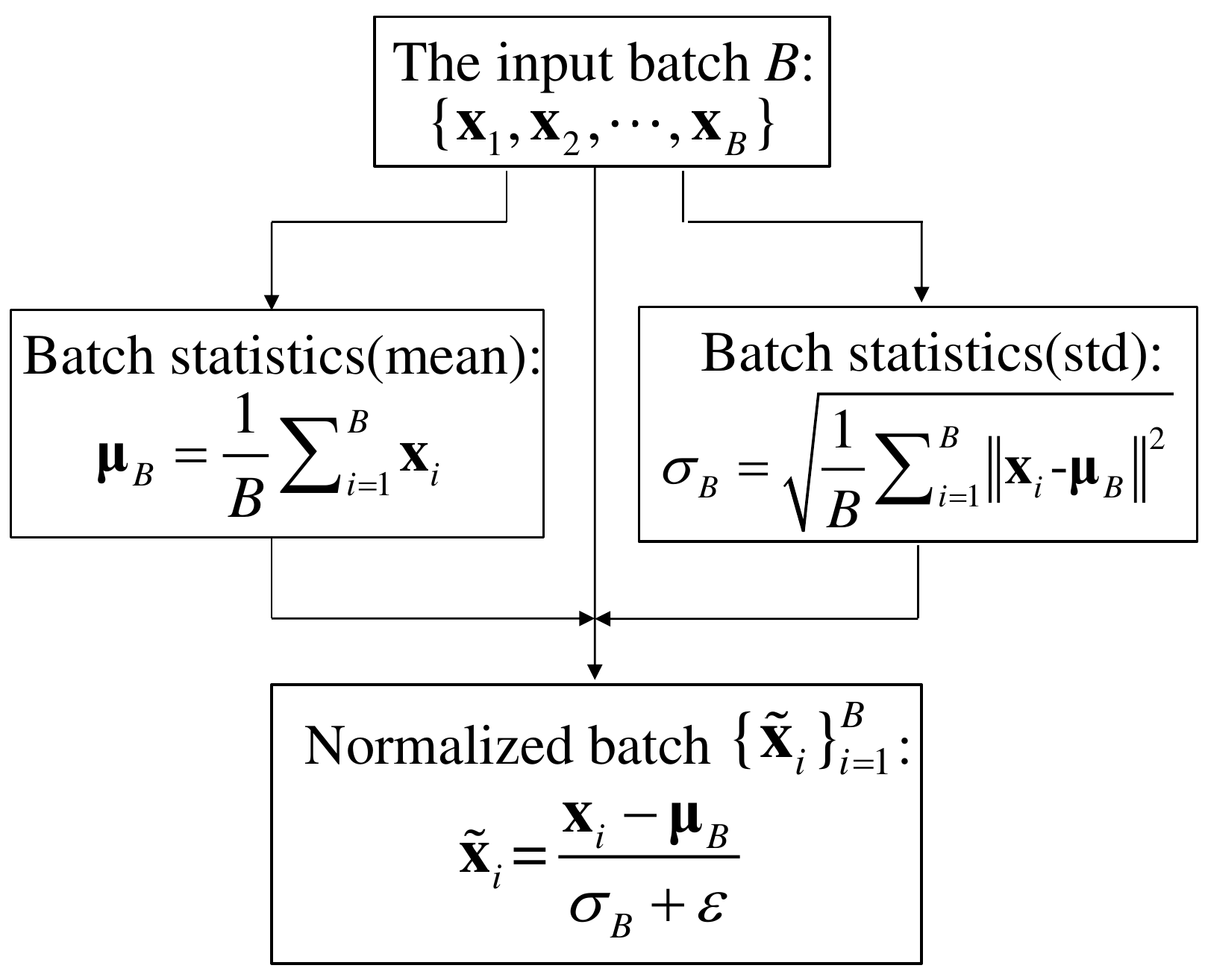}
     \caption{Batch Normalization}
   \end{subfigure}
   \begin{subfigure}{.24\textwidth}
      \centering
      \includegraphics[height=3.3cm, width=4cm]{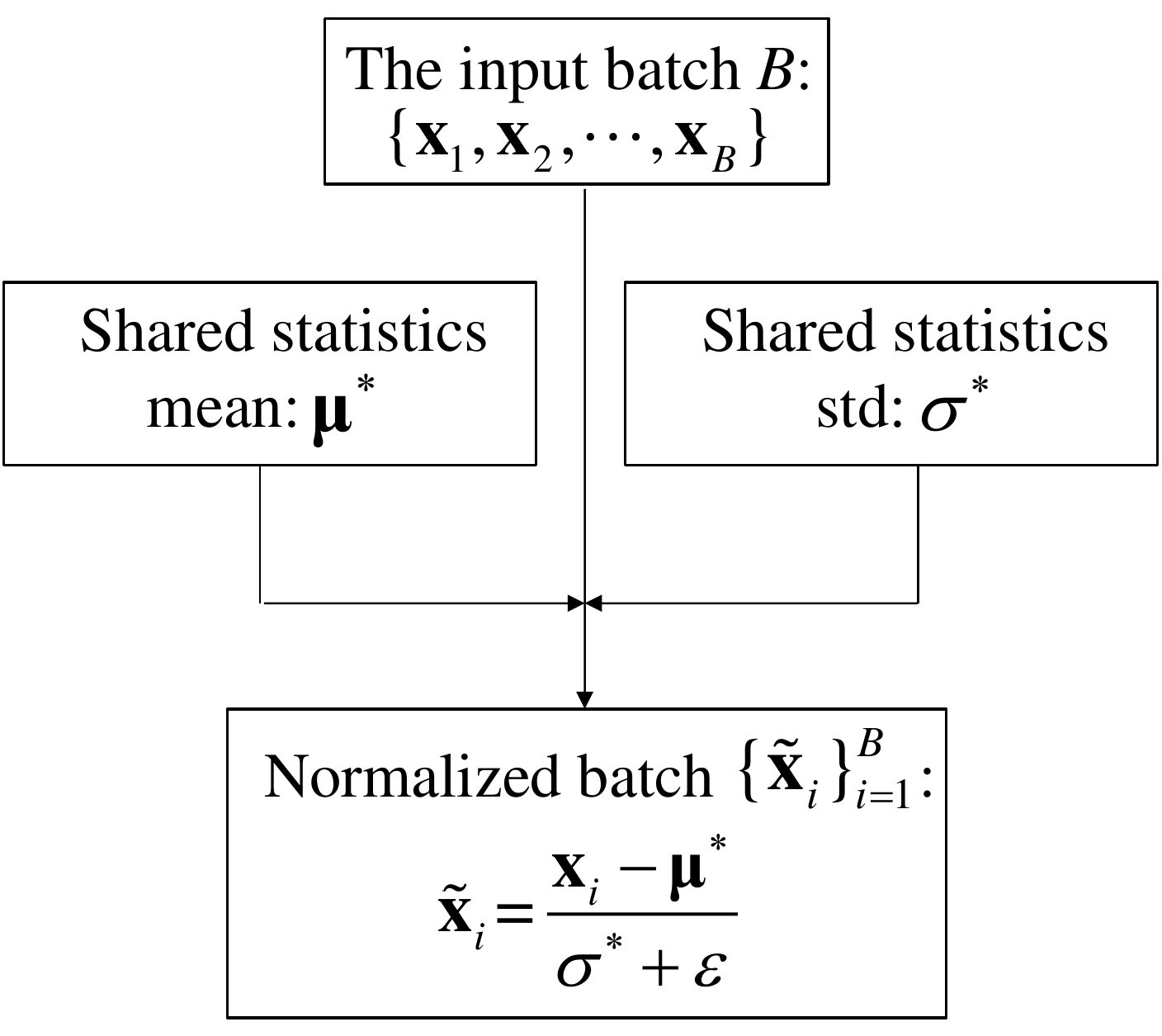}
      \caption{Shared Normalization}
   \end{subfigure}
   \caption{(a). The batch normalization layer normalizes the input batch with its own statistics. Scaling factor and bias factor are omitted in this figure. (b). The SN layer normalizes all input batches with the same statistics. }
\end{figure}

\begin{figure*}[t]
   \centering
   \includegraphics[height=3.3cm, width=18cm]{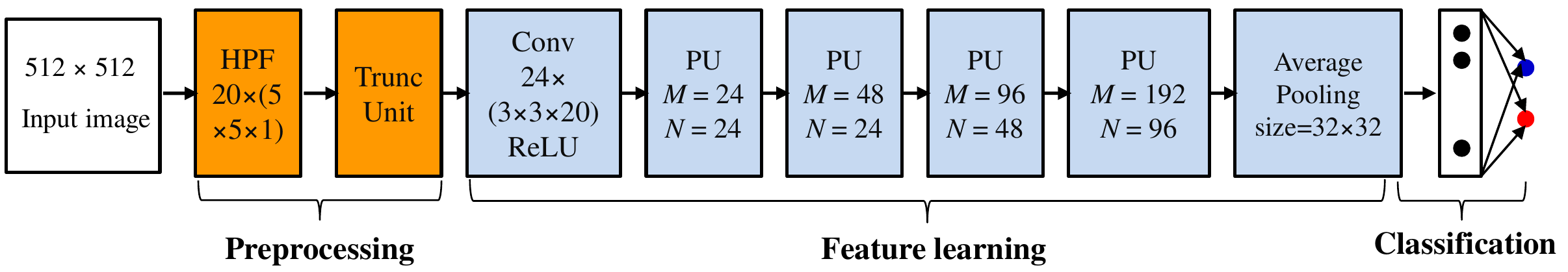}
      \caption{The proposed network for image steganalysis. The preprocessing subnetwork consists of the High Pass Filtering (HPF) layer and the truncation layer, where the HPF layer is to extract the noise component of input images and the truncation layer is to constrain the dynamic range of input feature map. The feature learning subnetwork contains several Processing Units (PU) to extract discriminative features for image steganalysis. The classification subnetwork maps extracted features into labels.}
\end{figure*}

Usually, batch staistics are calculated to normalize the input data in training phase but they are fixed in the testing phase. Our later experiment in Section V and the proof in Appendix A indicate that inaccurate estimation to the batch statistics in training would result in significant performance degradation in testing, which is similar to the analysis in unpaired case.

\section{A Novel Convolutional Neural Network for Image Steganalysis}
In this section, we first introduce a novel layer, called shared normalization layer. Then, based on this layer, a novel CNN model has been proposed for image steganalysis.
\subsection{Learning Features with Shared Normalization}
Previous analysis shows that a CNN model with multiple batch normalization layers is hard to be generalized to the test data when it is trained by paired learning and batch statistics are used to normalize the data. To address this difficulty, we propose to use commonly shared statistics rather than batch statistics to normalize input data in training and testing. Assume $\{(\mathbf{x}_{i}, \mathbf{y}_{i})\}_{i \in \mathcal{B}}$ denotes cover images and its stegos in a mini-batch $\mathcal{B}$. The proposed SN normalizes input data as:
\begin{equation}
  \mathbf{x}^{s}_{i} = SN(\mathbf{x}_{i}) = \frac{\mathbf{x}_{i}-\mu^{*}}{\sigma^{*}+\epsilon}
\end{equation}
\begin{equation}
  \mathbf{y}^{s}_{i} = SN(\mathbf{y}_{i}) = \frac{\mathbf{y}_{i}-\mu^{*}}{\sigma^{*}+\epsilon}
\end{equation}
where $\mu^{*}$ and $\sigma^{*}$ represents the estimated mean and standard deviation to all cover images and stego images, rather than the batch mean and batch standard deviation. $\epsilon$ is a small value to avoid the denominator to be zero. The difference between the batch normalization layer and the proposed SN layer for data normalization is shown as Fig.3. With the SN layer, we can obtain two advantages. On one hand, the proposed SN can obtain stable and consistent statistical properties from training set, avoid the limitation of batch normalization that cover images and their stegos are required to be paired in training and testings. One the other hand, with common statistics, a CNN model with SN layers can learn discriminative patterns between cover images and stego images, thus the generalization ability is improved.

\subsection{Network Architecture}
Based on the SN layer, we propose a novel CNN model for image steganalysis. Fig.4 shows the overall architecture of the proposed network. The network contains the preprocessing sub-network, the feature learning subnetwork and the classification subnetwork, which are introduced in the following parts.

\begin{figure}[t]
   \centering
   \includegraphics[height=3.2cm, width=6cm]{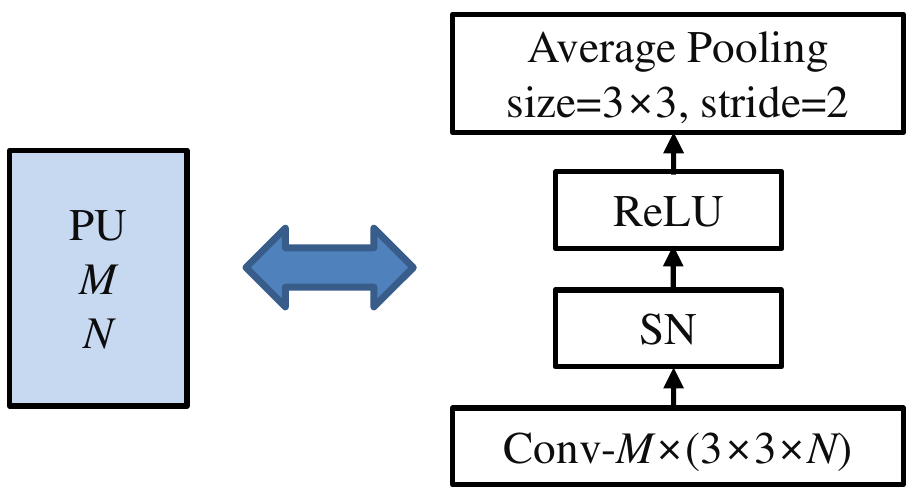}
      \caption{A Processing Unit (PU) contains a convolutional layer, a SN layer, a ReLU activation layer and a pooling layer. }
\end{figure}

The preprocessing subnetwork is to extract the high frequency component from input cover/stego images. It contains two layers, i.e. the High-Pass-Filtering (HPF) layer and the truncation layer. For HPF layer, it contains several highpass filters to extract high frequency signals from input images, aiming to remove most of image contents thus amplify the Signal-to-Noise Ratio(SNR) between stego signal and cover image. For those highpass filters, we follow the setting in [16] and use the 2nd order, 3rd order, EDGE-$3\times 3$, KV kernel and their rotations in our model. For the truncation layer, similar to [16-17], we use Eq.(26) to constrain the dynamic range of input feature map, in order to furtherly remove the image content and improve the convergence speed:
\begin{eqnarray}
  Tunc(x) =
   \left\{
   \begin{array}{lll}
      -T, \ x < -T \\
      x, \ \  -T \leq x \leq T \\
      T, \ \  x > T
   \end{array}
   \right.
\end{eqnarray}
where $T$ denotes the truncation threshold.

The feature learning subnetwork is to extract effective features for image steganalysis. In the feature learning subnetwork, 24 convolutional kernels with the size of $3 \times 3$ are used to process the the noise components generated by the preprocessing subnetwork. Following the convolutional layer is a ReLU activation layer [39]. Then, the network uses the Processing Unit (PU), which is depicted by Fig.5,  to process the coming data. A PU consists of a convolutional layer, a SN layer, a ReLU layer and an average pooling layer. The size of convolutional kernels in the block is $3 \times 3$ and the number of convolutional kernel is 24, 48, 96 and 192. To make the network learn discriminative features for steganalysis consistently, we force that the size of pooling layer is same to the convolutional kernel, i.e. $3 \times 3$. After several PUs, an average pooling layer with large size, $32 \times 32$, transforms feature maps into feature vectors. The purpose of using a large size pooling layer is to reduce the information loss introduced by many processing layers.

The classification subnetwork maps extracted features into binary labels. Two output nodes, which corresponds to the label of cover and stego, are fully connected to the feature map that are averaged pooled in the feature learning subnetwork.

\subsection{Parameter Learning}
Parameters of the proposed network are learned by minimizing the softmax loss function:
\begin{equation}
   L(\mathbf{x_{i}}, \mathbf{\theta} ) = -\sum_{k=1}^{K} 1\{y_{i} = k\} \cdot log \left( \frac{e^{o_{i,k}(\mathbf{x_{i}}, \mathbf{\theta} ) } } { \sum_{k=1}^{K} e^{o_{i,k}(\mathbf{x_{i}}, \mathbf{\theta} ) }} \right)
\end{equation}
where $\mathbf{\theta}$ denotes the parameters of the network, including weight matrices $\mathbf{W}$ and the bias vectors $\mathbf{b}$. $K$ is the number of labels, where $K=2$ in our model. $y_{i}$ is the label of $\mathbf{x}_{i}$, $1\{\cdot\}$ is the indicator function. $o_{i,k}(\mathbf{x_{i}}, \mathbf{\theta} )$ represents the output of the network for the sample $\mathbf{x}_{i}$. $\mathbf{W}$ and $\mathbf{b}$ of the network parameter $\mathbf{\theta}$ are updated by the mini-batch stochastic gradient descending (SGD):
\begin{equation}
  \mathbf{W}(t+1) = \mathbf{W}(t) - \alpha_{1} \frac{1}{N} \sum_{i \in \mathcal{B}} \frac{\partial L(\mathbf{x}_{i},\mathbf{\theta})}{\partial \mathbf{W} }
\end{equation}
\begin{equation}
  \mathbf{b}(t+1) = \mathbf{b}(t) - \alpha_{1} \frac{1}{N} \sum_{i \in \mathcal{B}} \frac{\partial L(\mathbf{x}_{i},\mathbf{\theta})}{\partial \mathbf{b} }
\end{equation}
where $N$ is the size of a mini-batch $\mathcal{B}$, $\alpha_{1}$ is the learning rate.

{\setlength{\belowcaptionskip}{-10pt}
  \begin{figure}[t]
   \includegraphics[width=2cm,height=2cm]{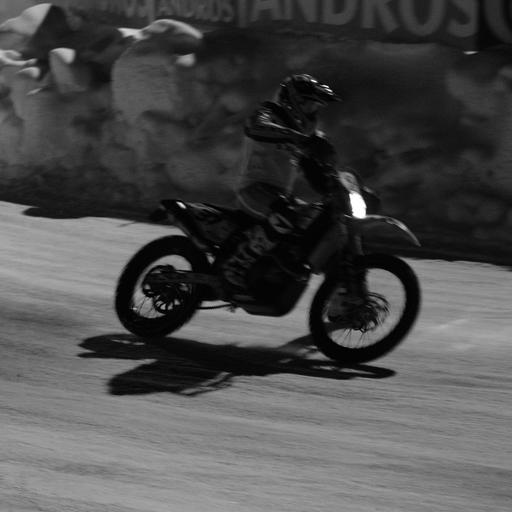}
   \includegraphics[width=2cm,height=2cm]{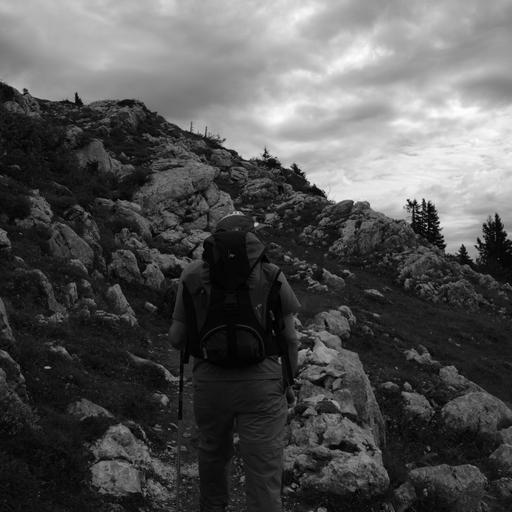}
   \includegraphics[width=2cm,height=2cm]{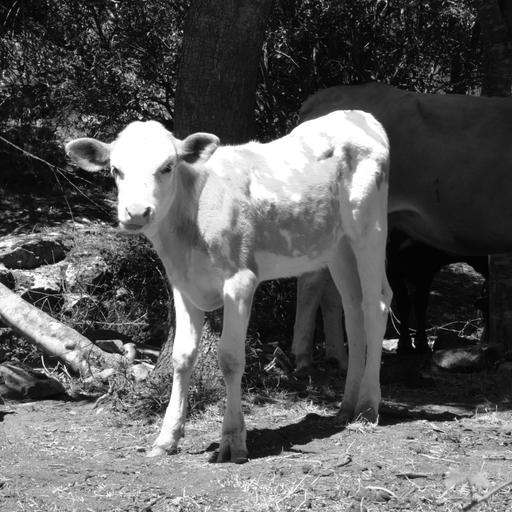}
   \includegraphics[width=2cm,height=2cm]{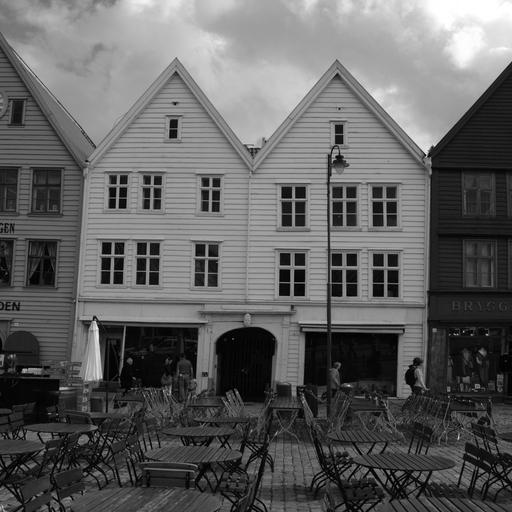}
   \centering
   \caption{Sample images in BOSSbase ver 1.01.}
 \end{figure}}

\begin{figure*}[t]
   \centering
   \begin{subfigure}{.45\textwidth}
     \centering
     \includegraphics[height=6.5cm, width=8.5cm]{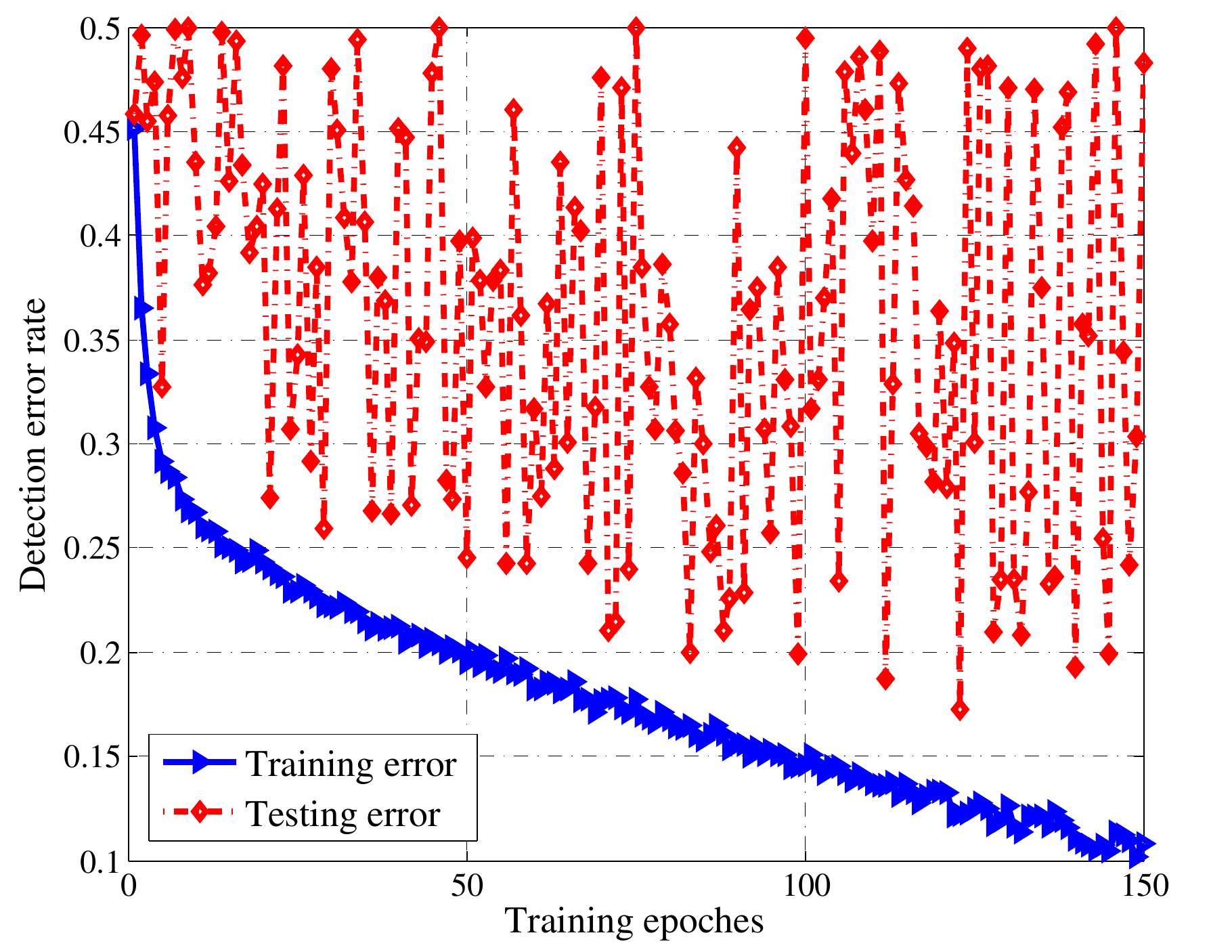}
     \caption{Baseline network with batch normalization layers}
   \end{subfigure}
   \begin{subfigure}{.45\textwidth}
      \centering
      \includegraphics[height=6.5cm, width=8.5cm]{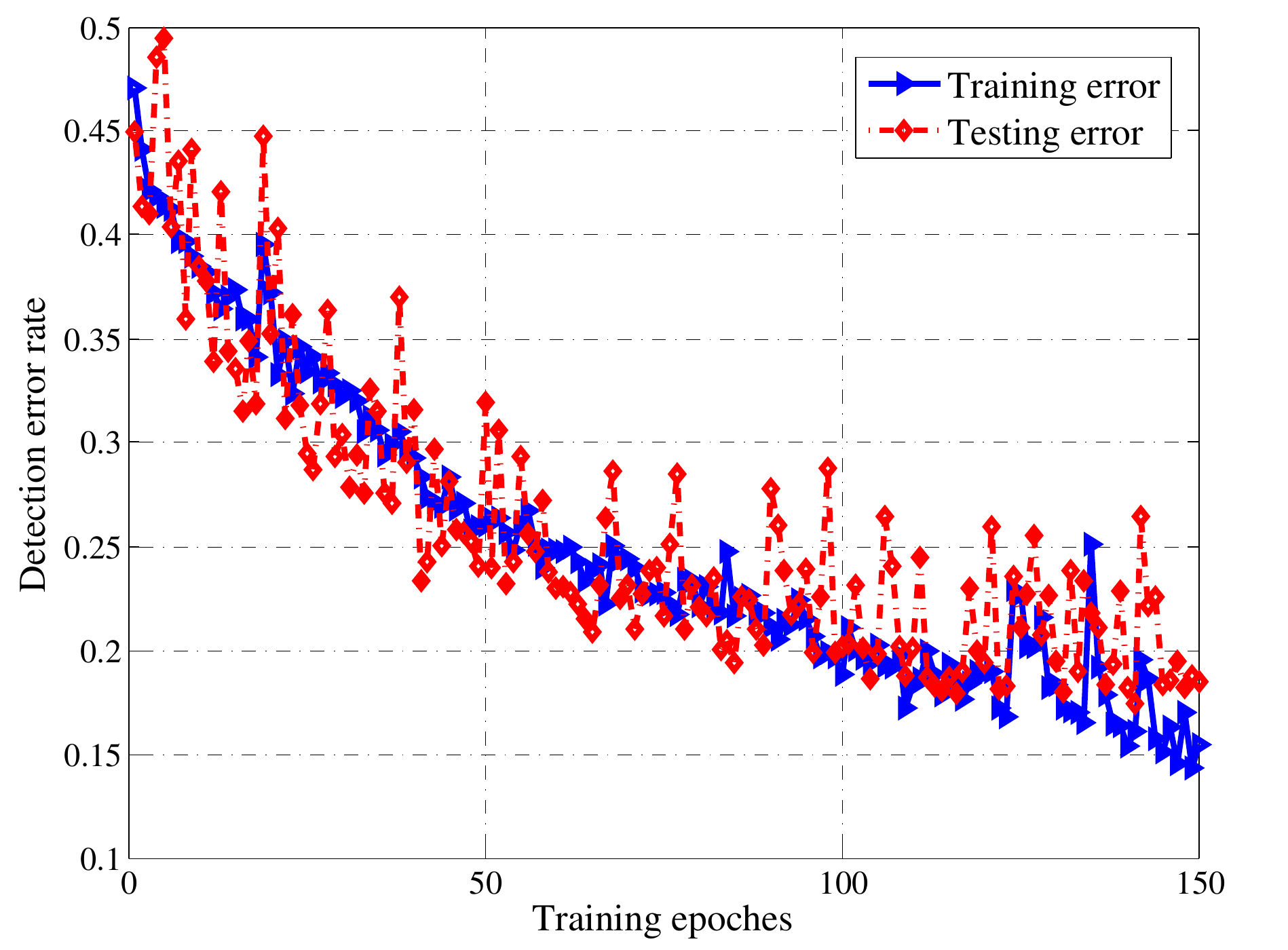}
      \caption{Proposed network with SN layers}
   \end{subfigure}
   \caption{Training error curve and testing error curve of the baseline network and the proposed network in the first 150 training epoches. The S-UNIWARD steganography at payload 0.4 bpp is used for demonstration. }
\end{figure*}

\section{Experiments}
In this section, comprehensive experiments are conducted to demonstrate the effectiveness of the proposed SN layer and our new CNN model for image steganalysis. We not only compare the proposed network with the state-of-the-art rich model method but also compare it with several recently proposed CNN models. We further conduct sensitivity experiments to test our model when the embedding payload or steganographic algorithm is mismatched. We finally demonstrate that some techniques widely used in deep learning, including data augmentation and model ensemble, can also improve our CNN model's detection accuracy.

\subsection{Dataset and Steganographic Schemes}
The dataset used for validation is the BOSSbase 1.01 [28], which is a standard database for evaluating steganography and steganalysis. The BOSSbase contains 10,000 uncompressed natural images with the size of $512 \times 512$, including human beings, natural scenes, animals and buildings, which are shown as Fig.6. For performance evaluation, the detection error rate $P_{E}$ is used to measure the security of different steganographic algorithms:
 \begin{equation}
   P_{E} = \frac{1}{2}(P_{MD}+P_{FA})
 \end{equation}
where $P_{MD}$ is the miss detection probability and $P_{FA}$ represents the false alert probability.

Four states of the art steganographic algorithms, including the Highly Undetectable steGanOgraphy with Bounding Distortion (HUGO-BD) [34], the Wavelet Obtained Weights steganography (WOW) [35], S-UNIWARD [33], and the HIgh-pass Low-pass Low-pass steganography (HILL) [36], are used for performance validation. To avoid that secret data are embedded with the same key, we adopt the MATLAB implementation of steganographic algorithms rather than the C++ version for performance evaluation. In all experiments, we use bit-per-pixel (bpp) to represent the size of secret data embedded into cover images.

{\setlength{\abovecaptionskip}{2pt}
 \setlength{\belowcaptionskip}{-2pt}
\begin{table*}[t]
  \centering
  \renewcommand\arraystretch{1.2}
  \caption{Detection error rates of SRM, maxSRM and the proposed network for four steganographic algorithms at five different payloads. The BOSSbase dataset is used for validation. }
  \resizebox{14.5cm}{!} {
  \begin{tabular}{| c | c | c | c | c | c | c | }
  \hline
    \textbf{Steganography} & \textbf{Detection algorithm} & 0.05 bpp & 0.1 bpp & 0.2 bpp & 0.3 bpp & 0.4 bpp \\ \hline
    \multirow{3}{*}{HUGO-BD} & SRM + ensemble & 42.60\% & 37.26\% & 28.78\% & 22.54\% & 18.23\% \\
                         & maxSRM + ensemble & 36.83\% & 31.32\% & 24.53\% & 20.37\% & 16.47\% \\
                         & The proposed network & \textbf{36.76\%} & \textbf{30.81\%} & \textbf{23.72\%} & \textbf{19.25\%} & \textbf{15.43}\% \\ \hline
    \multirow{3}{*}{WOW} & SRM + ensemble & 45.63\% & 40.15\% & 32.31\% & 25.66\% & 20.08\%  \\
                         & maxSRM + ensemble & \textbf{35.39\%} & 30.18\% & 23.84\% & 18.92\% & 15.40\% \\
                         & The proposed network & 35.87\% & \textbf{30.02\%} & \textbf{23.48\%} & \textbf{17.43\%} & \textbf{14.26\%} \\ \hline
    \multirow{3}{*}{S-UNIWARD} & SRM + ensemble & 45.38\% & 40.38\% & 32.54\% & 25.51\% & 20.70\%\\
                         & maxSRM + ensemble & \textbf{41.98\%} & 35.63\% & 28.04\% & 22.35\% & 18.84\% \\
                         & The proposed network & 42.13\% & \textbf{35.21\%} & \textbf{26.82\%} & \textbf{20.71\%} & \textbf{16.53\%} \\ \hline
    \multirow{3}{*}{HILL} & SRM + ensemble & 47.32\% & 43.71\% & 36.47\% & 29.39\% & 24.57\% \\
                         & maxSRM + ensemble & 42.31\% & 37.76\% & 30.95\% & 25.71\% & 21.63\% \\
                         & The proposed network & \textbf{42.15\%}  & \textbf{36.86\%} & \textbf{29.63\%} & \textbf{23.60\%} & \textbf{19.87\%} \\ \hline
  \end{tabular}
  }
\end{table*}}

\subsection{Parameter Setting and Implementation Details}
For the proposed network, each element $W_{ij}$ in the weight matrix $\mathbf{W}$ is initialized by the improved ``Xavier" method [37], i.e. Gaussian distribution with zero mean and the standard deviation inversely proportional to the number of network's connections:
\begin{equation}
   W_{ij} \sim \mathcal{N}\left(0,\frac{2}{c_{n}}\right)
\end{equation}
where $c_{n}$ denotes the number of connections at $n-$th layer. The momentum and the weight decay in our network is set to 0.9 and 0.0001 respectively.  The threshold $T$ in the truncation layer is set to 5. The size of mini-batch SGD, $N$, is set to 40, including 20 cover images and 20 corresponding stego images. The number of training epoch is set to 200. Similar to existing deep CNN models [15], instead of using a fixed learning rate, we adaptively adjust it in training phase. Specifically, the learning rate $\alpha_{1}$ for $\mathbf{W}$ is set to 0.01 in first 150 training epoches and is decreased to 0.001 in resting 50 epoches. The purpose of decreasing the learning rate is to make the network escape the error plateaus.

For the SN layer, the mean $\mathbf{\mu}_{n}^{*}$ and the standard deviation $\sigma_{n}^{*} $ for each feature map are initialized by:
\begin{equation}
   \mu_{n}^{*} = \frac{1}{M}\sum_{i=1}^{M} E\left[ vec(\mathbf{f}^{i}_{n-1}) \right]
\end{equation}
\begin{equation}
   \sigma_{n}^{*} = \frac{1}{M}\sum_{i=1}^{M} Var\left[ vec(\mathbf{f}^{i}_{n-1}) - \mu_{n}^{*}  \right]
\end{equation}
where $vec(\cdot)$ represents the vectorization operation, $\mathbf{f}^{i}_{n-1}$ denotes the feature map for the $i$-th input sample at previous layer. To make the $\mu_{n}^{*}$ and $\sigma_{n}^{*}$ be accurate estimations to the mean and standard deviation of the whole dataset, we set $M$ to 200, including 100 cover images and 100 their stego versions, to initialize $\mu_{n}^{*}$ and $\sigma_{n}^{*}$ in the SN layer. During training, $\mu_{n}^{*}$ and $\sigma_{n}^{*}$ are not fixed but are updated by following equations:
\begin{equation}
  \mu_{n}^{*} (t+1)  = (1-\alpha_{2})\mu_{n}^{*} (t) + \alpha_{2}\mu_{n}^{\mathcal{B}}(t)
\end{equation}
\begin{equation}
  \sigma_{n}^{*} (t+1)  = (1-\alpha_{2})\sigma_{n}^{*} (t) + \alpha_{2}\sigma_{n}^{\mathcal{B}}(t)
\end{equation}
where $\alpha_{2}$ is the learning rate for SN, $\mu_{n}^{\mathcal{B}}(t)$ and $\sigma_{n}^{\mathcal{B}}(t)$ are the mean and standard deviation calculated by the mini-batch $\mathcal{B}$ at $t$-th iteration. To make SN layers capture the statistics of the network, $\alpha_{2}$ are set same to $\alpha_{1}$, i.e. 0.01 in first 150 epoches and 0.001 in the resting 50 epoches.

\begin{figure*}[t]
   \centering
   \begin{subfigure}{.45\textwidth}
     \centering
     \includegraphics[height=6.7cm, width=8cm]{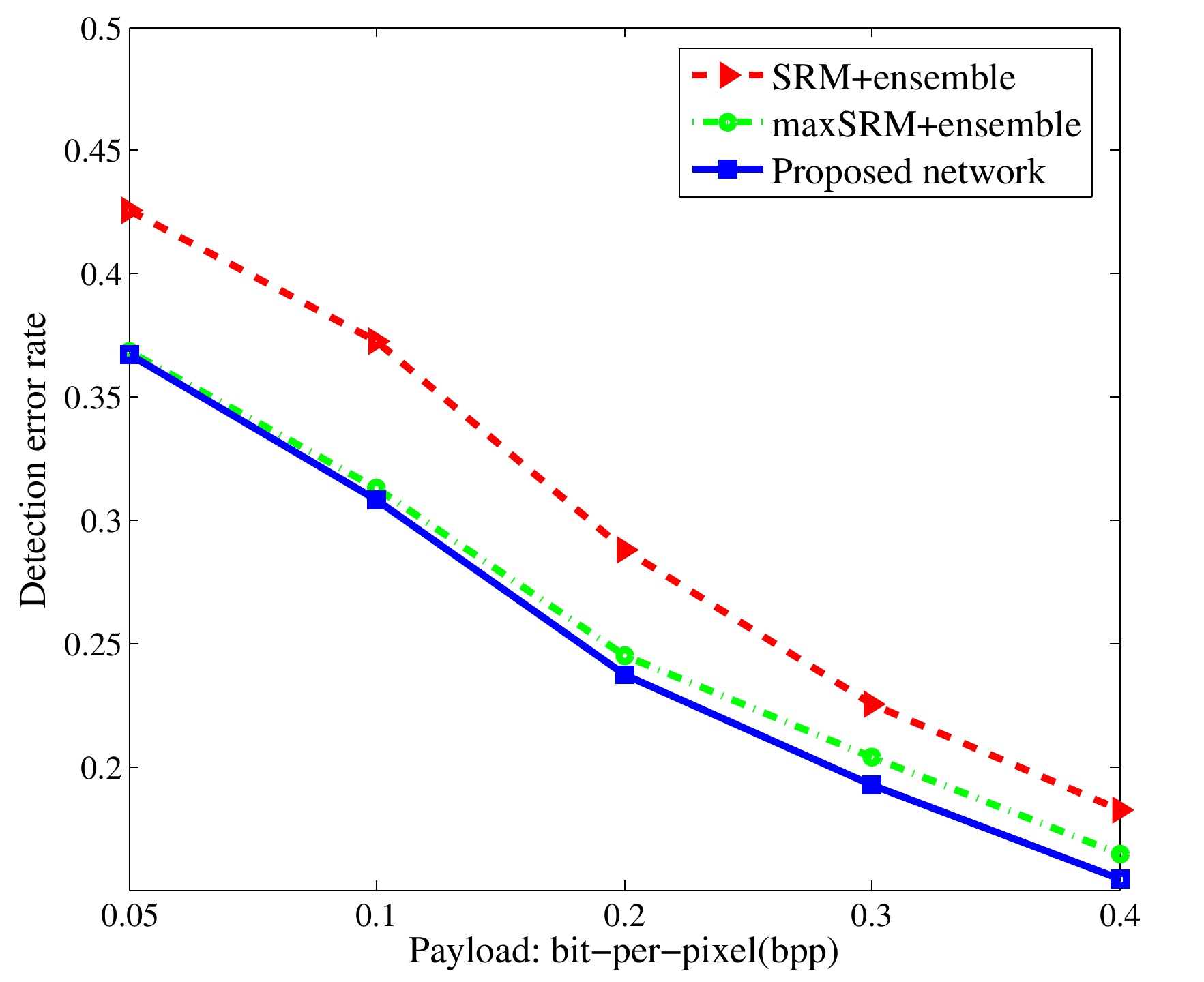}
     \caption{HUGO-BD}
   \end{subfigure}
   \begin{subfigure}{.45\textwidth}
      \centering
      \includegraphics[height=6.7cm, width=8cm]{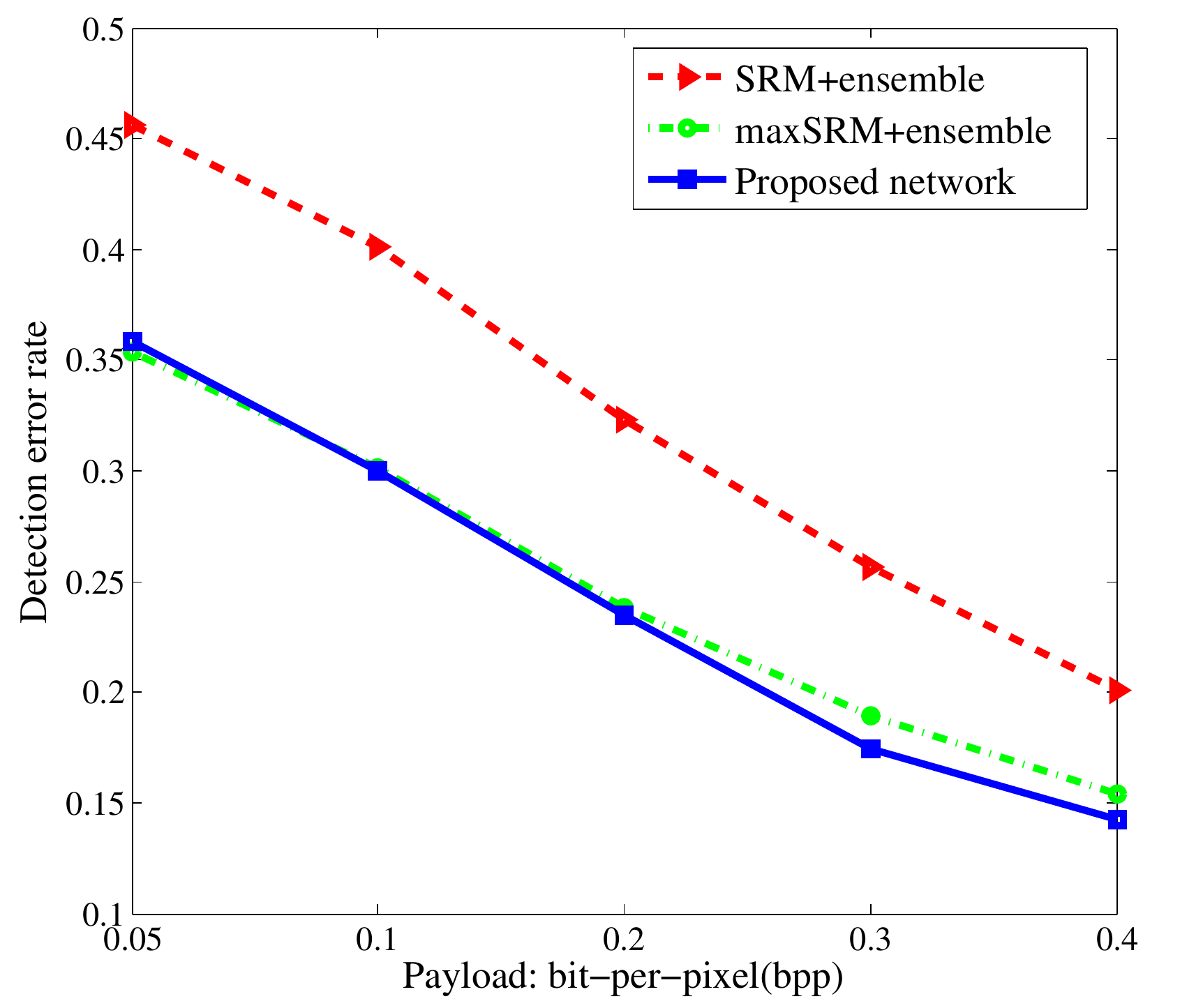}
      \caption{WOW}
   \end{subfigure}
   \begin{subfigure}{.45\textwidth}
     \centering
     \includegraphics[height=6.7cm, width=8cm]{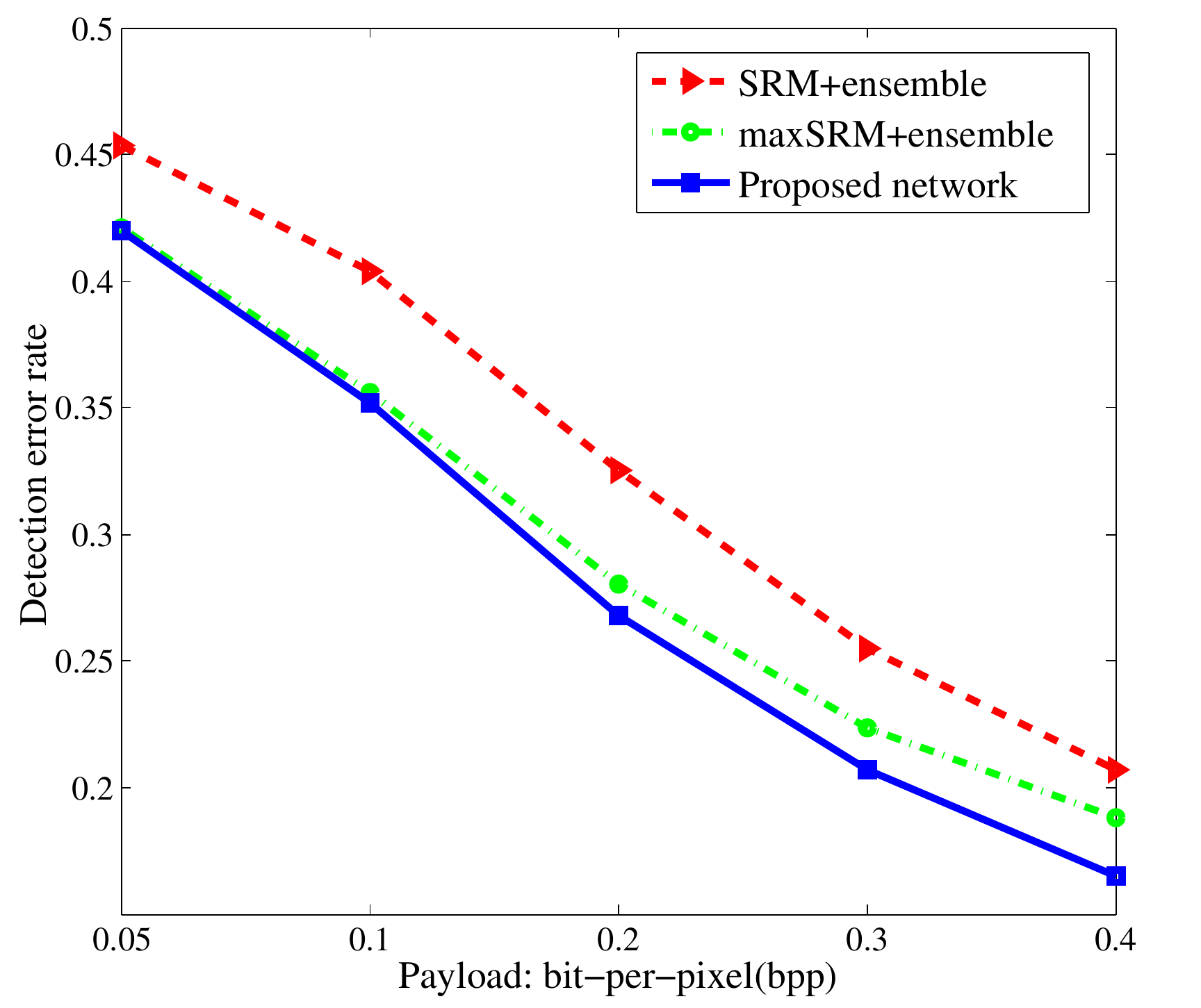}
     \caption{S-UNIWARD}
   \end{subfigure}
   \begin{subfigure}{.45\textwidth}
      \centering
      \includegraphics[height=6.7cm, width=8cm]{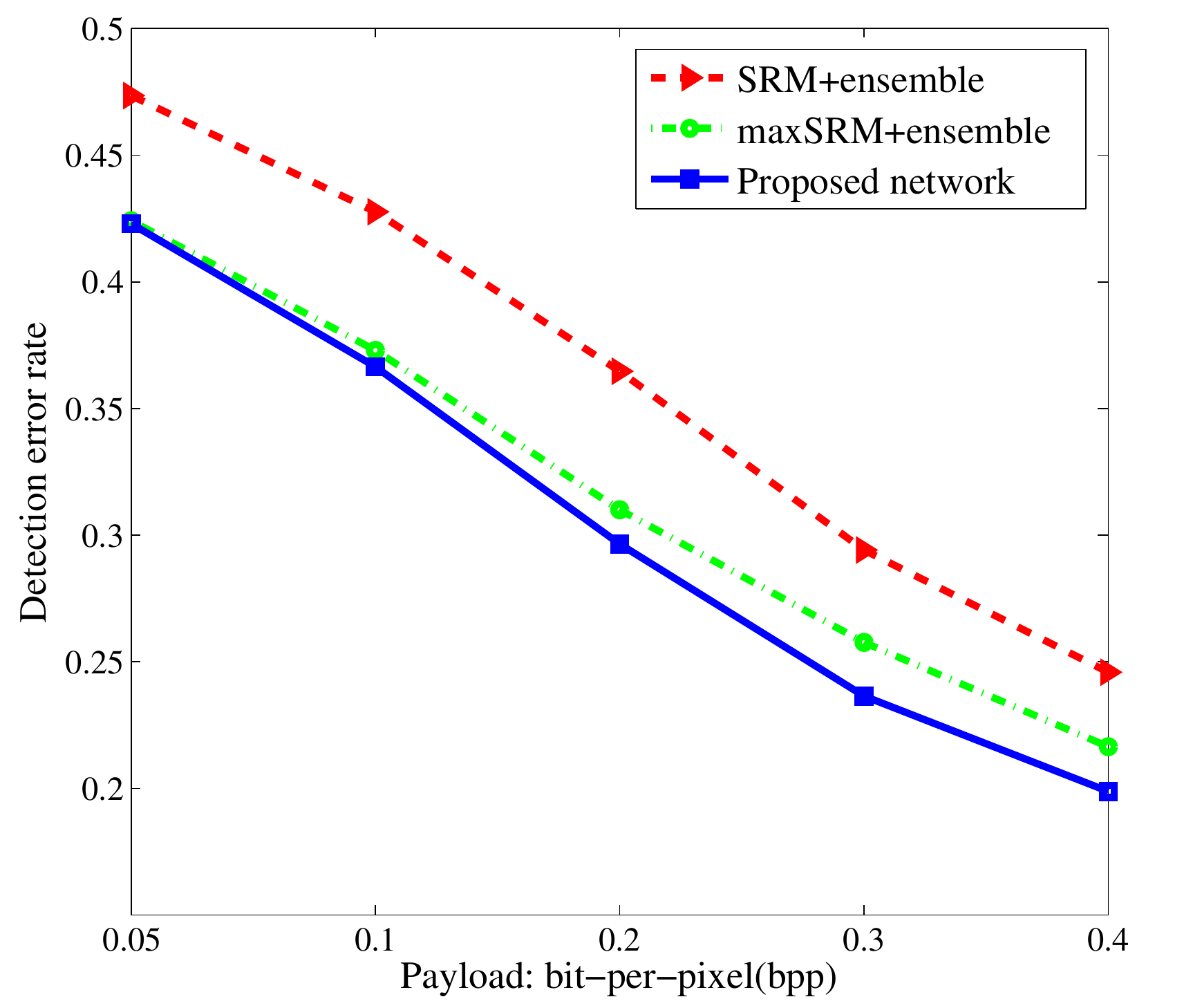}
      \caption{HILL}
   \end{subfigure}
   \caption{Graphical representation of detection error rates for SRM, maxSRM and the proposed network. (a) HUGO-BD steganography. (b) WOW steganography. (c) S-UNIWARD steganography. (d) HILL steganography. }
\end{figure*}

\begin{figure*}[t]
   \centering
   \begin{subfigure}{.3\textwidth}
     \centering
     \includegraphics[height=5cm, width=5cm]{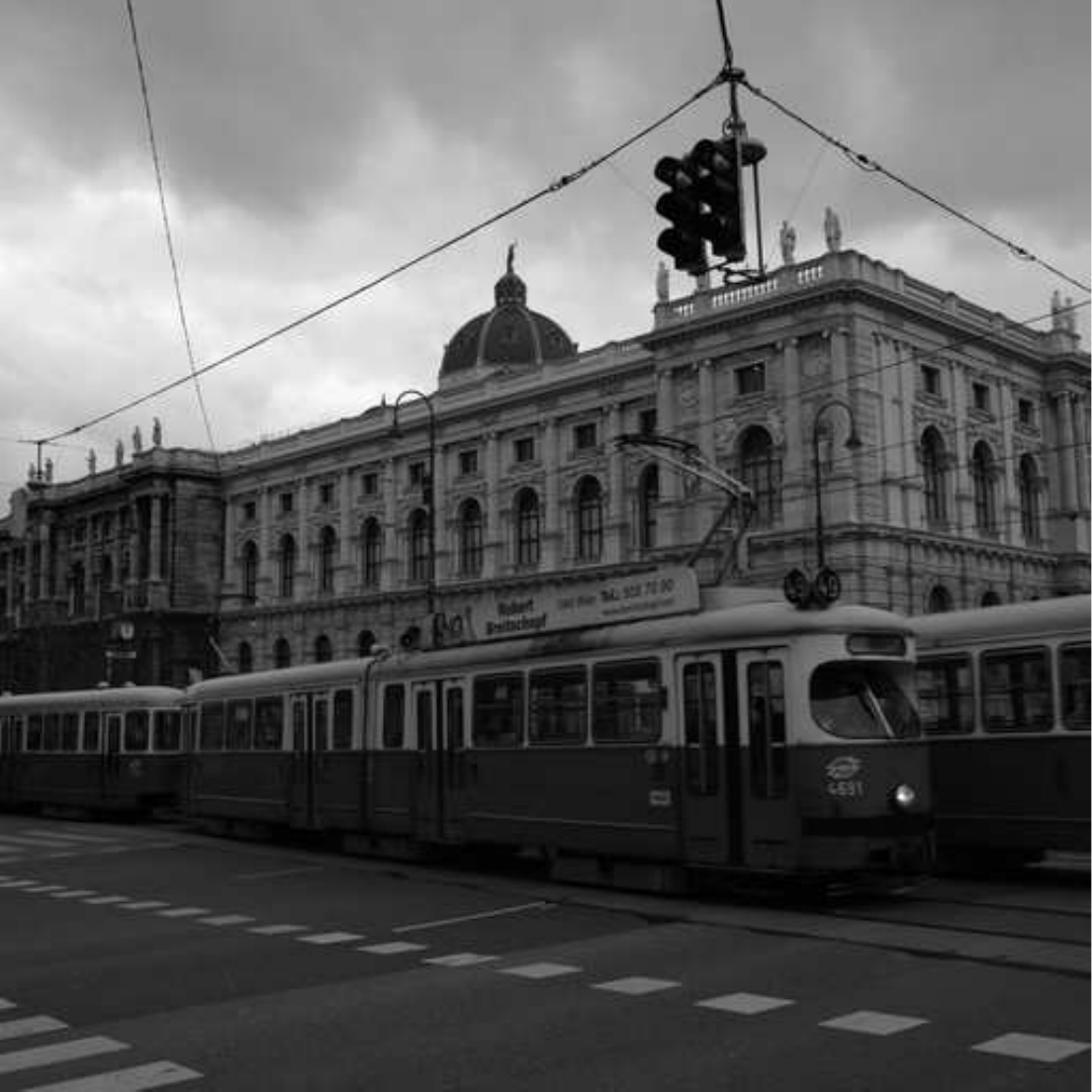}
     \caption{Cover image}
   \end{subfigure}
   \begin{subfigure}{.3\textwidth}
      \centering
      \includegraphics[height=5cm, width=5cm]{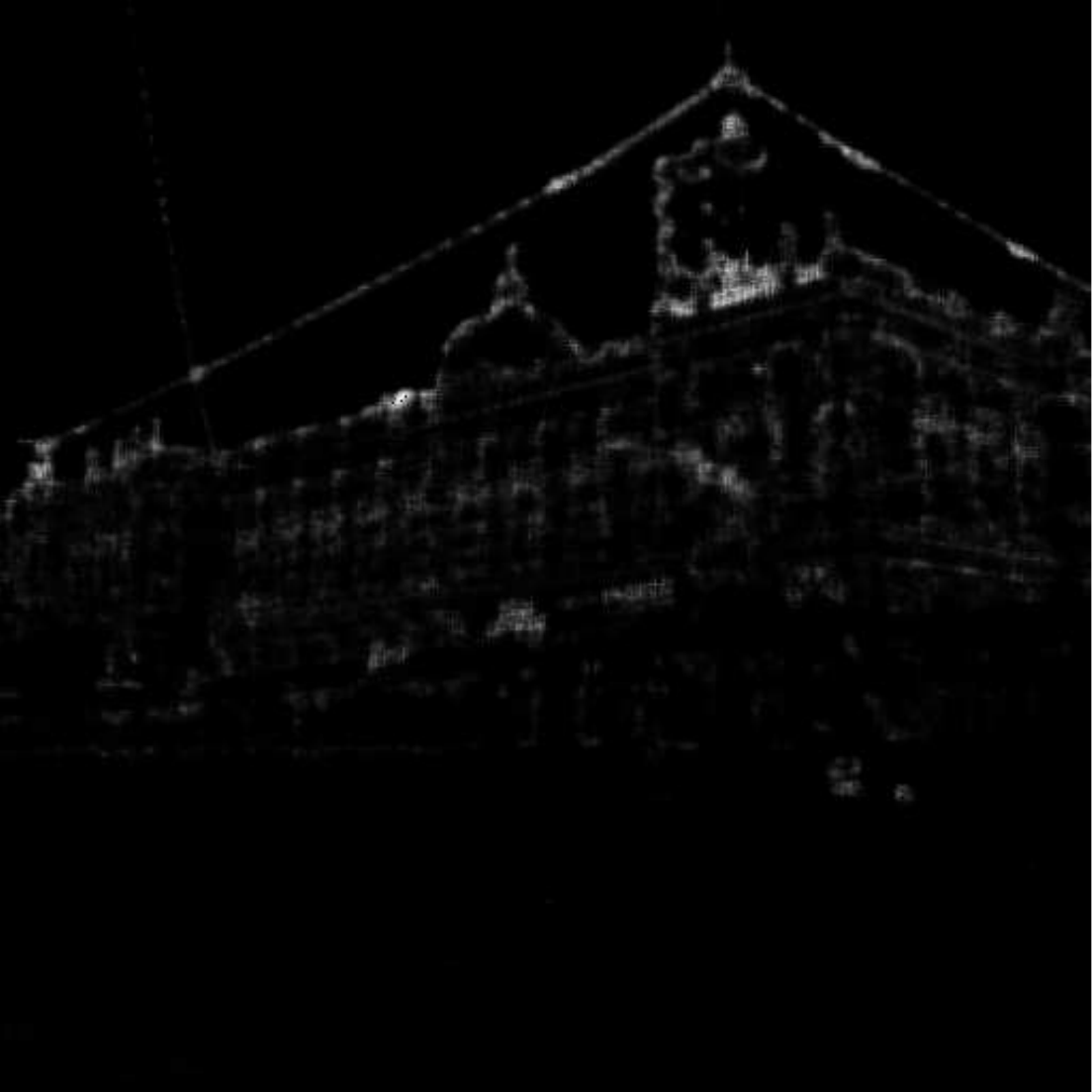}
      \caption{Embedding probability map at 0.05 bpp}
   \end{subfigure}
   \begin{subfigure}{.3\textwidth}
      \centering
      \includegraphics[height=5cm, width=5cm]{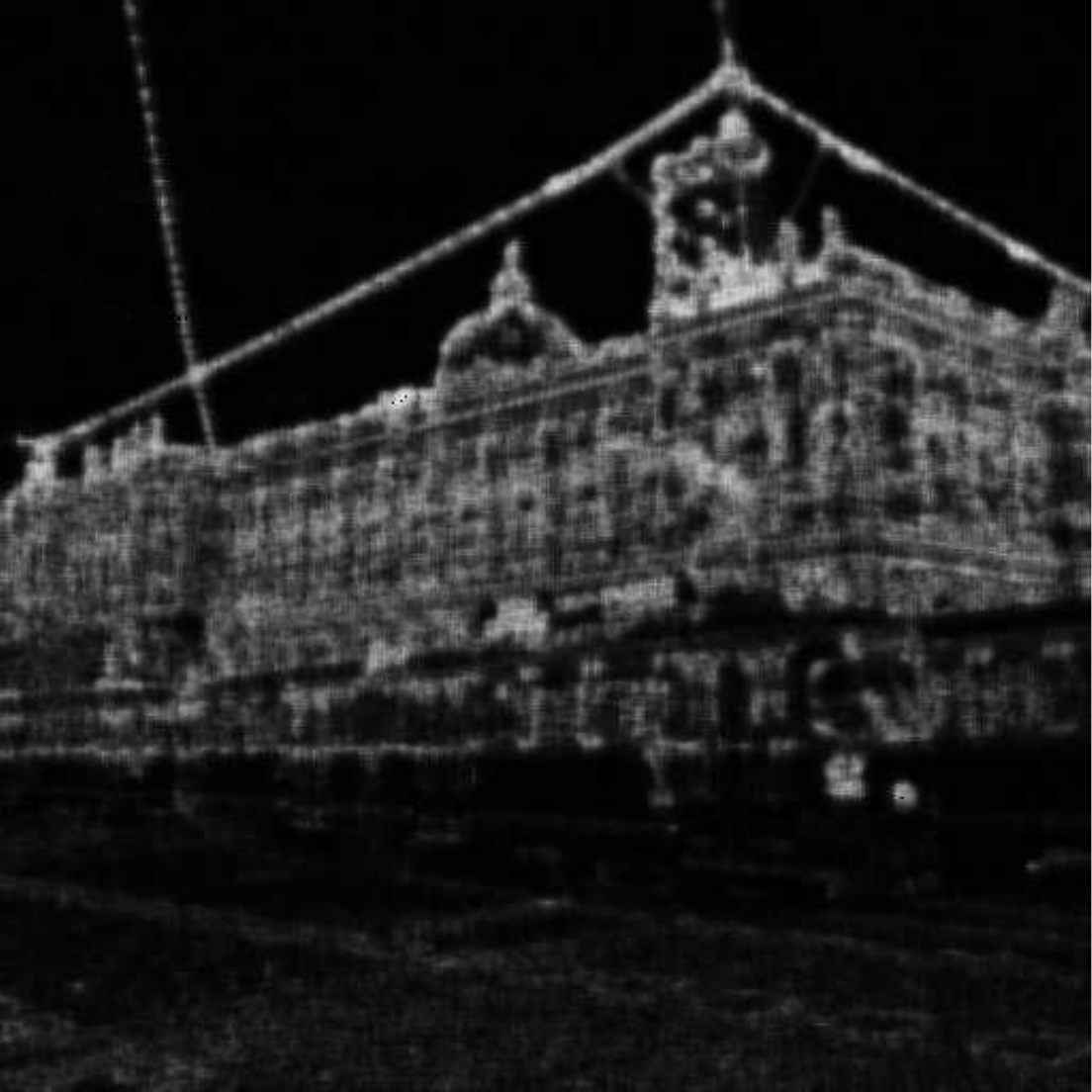}
   \caption{Embedding probability map at 0.4 bpp}
   \end{subfigure}
   \caption{A cover image and its embedding probability maps at different payloads. The S-UNIWARD steganography at 0.05 bpp and 0.4 bpp are used for demonstration. (a). the ``433.pgm" cover image in the BOSSbase dataset; (b). the embedding probability at 0.05 bpp; (c). the embedding probability at 0.4 bpp. }
\end{figure*}

\subsection{Comparison: Network with Batch Normalization Layers and Network with Shared Normalization Layers}
In this experiment, we demonstrate the effectiveness of the proposed shared normalization statistics layer over the batch normalization layer for image steganalysis. The proposed network with SN layers is compared with a baseline network. To make the result comparable, the baseline network has exactly same architecture with the proposed network except that all SN layers are replaced with batch normalization layers. In addition, two networks utilize same training images and testing images in the experiment. The S-UNIWARD stegnography at the payload 0.4 bpp is used for demonstration.

Fig.7 shows the training error and testing error of the baseline network and the proposed network. For the baseline network, the training error decreases fast as the number of epoch increases. However, the testing error vibrates greatly and no clear connection between the training error and the testing error can be found during the training phase. This indicates that features learned by the baseline network can hardly contribute to the classification of cover images and stego images. For the proposed network, the testing error systematically decreases as the training error decreases, demonstrating that the proposed network with SN layers can effectively learn discriminative features to classify unknown cover images and stego images.

\subsection{Performance Comparison with Prior Arts}
In this subsection, we conduct comprehensive experiments to demonstrate the effectiveness of the proposed CNN model. We compare the proposed network with the classical SRM [5] and its select-channel-aware version, the maxSRM steganalysis [6]. SRM steganalysis extracts many handcrafted features that are sensitive to message embedding and combine them into a long feature vector for classification. An ensemble classifier [30] is trained based on the extracted features and is used for predicting the label of an input image. With embedding probability maps, the maxSRM steganalysis focuses more on image pixels with lower embedding distortions. Four steganographic algorithms, i.e. HUGO-BD, WOW, S-UNIWARD, and HILL, at five different payloads, i.e. $[0.05, 0.1, 0.2, 0.3, 0.4]$, are used for validation. In the experiment, we randomly select 5,000 cover images and their stego images to train an optimized model, while the rest 5,000 cover images and their stego images are used for testing.

For network training, we use transfer learning [9, 16-17] to learn effective features for image steganalysis at low payloads. This technique is widely used in many deep convolutional neural network models [38-39], showing promising performances for closely related tasks. Specifically, our CNN model for a lower payload steganalysis, e.g. 0.3 bpp, are finetuned based on the network trained at a higher payload, e.g. 0.4 bpp. The reason is that directly learning discriminative features proves to be hard for CNN based steganalysis at low payloads [17], while transfer learning can regularize the feature space and utilize auxiliary information from stego images at higher payloads [16]. To avoid training samples are reused for testing at different payloads, we force that cover images for network training/testing at a lower payload are same to those for network training/testing at a higher payload.

TABLE 1 gives performance comparisons between the proposed network and SRM, maxSRM at different configurations. Compared with SRM steganalysis, the proposed network shows obvious performance improvements on all steganographic schemes at all payloads. For the maxSRM steganalysis, our network also shows better performances in most of cases even though no extra information, i.e. embedding probability map, is provided. Each element in an embedding probability map denotes the probability that a secret message is embedded at this position. From the graphical representation of the table, i.e. the Fig.8,  we have an important observation that the performance improvement between the proposed network and maxSRM becomes smaller as the payload decreases. This phenomenon has also been observed in [17], indicating that CNN models are not effective to detect steagnography at low payloads even though transfer learning is used. To illustrate this phenomenon, two embedding probability maps at extreme payloads, i.e. the largest payload 0.4 bpp and the smallest payload 0.05 bpp, are drawn as Fig.9. Unlike the embedding probability map at 0.4 bpp (as Fig.9(c) shows) that secret data can be almost embedded in the whole building area, the map at 0.05 bpp (as Fig.9(b) shows) only hides secret data in very complex regions such as sharp edges or textured patterns. Consequently, we conclude that two reasons may limit the performance of CNN models to detect steganography at low payloads. For the first, these very complex regions do not take a large proportion in most of natural images. In this case, CNN models become easily overfitted because they do not have enough training data to model the distribution of complex regions. For the second, very complex regions often show some properties like random ``noise" [40]. CNN models are proved to have strong ability extract regular patterns among natural images, but may not be effective to model this type of noisy patterns.

{\setlength{\abovecaptionskip}{2pt}
 \setlength{\belowcaptionskip}{-2pt}
\begin{table}[t]
  \centering
  \renewcommand\arraystretch{1.2}
  \caption{Detection error rates for CNN models on three steganographic algorithms at payload 0.4 bpp. `$-$' denotes that the result is not reported in the paper. }
  \resizebox{8.7cm}{!} {
  \begin{tabular}{| c | c | c | c |}
   \hline
  \textbf{CNN Models} & WOW & S-UNIWARD & HILL \\ \hline
  Qian's network [9] & 21.95\% & 22.05\% & $-$ \\ \hline
  Xu's network [10] & $-$ & 20.97\% & 22.42\% \\ \hline
  Proposed network & \textbf{14.26\%} & \textbf{16.53\%} & \textbf{19.87\%} \\ \hline
  \end{tabular}
  }
\end{table}}

Additionally, we compare the proposed network with two states of the art CNN based steganalytic methods, including Xu's network [10] and Qian's network [9]. For Xu's network, we report the detection error rates of a single model rather than the results by ensembling several models in [10]. For Qian's network, we report the lowest detection error rates for different steganography since the model has been transferred with several different payloads. TABLE II gives the performance comparison for three CNN models. It is observed that the proposed network also achieves much lower detection error rates on different steganographic schemes.

\subsection{Sensitivity Experiments}
Most existing work including our previous experiment assume that the data embedding algorithm and the payload are known to the steganalyzer.  However, it is often hard to obtain such information in real applications. For steganalytic algorithms, they are often difficult to detect steganography when the payload or the embedding algorithm is mismatched. Here, ``mismatch" means that the payload or the embedding algorithm used to train the model is different from that is used to test the model. This reality motivates us to investigate whether the proposed network is sensitive to the payload mismatch and the embedding algorithm mismatch. We use the following experiments to validate our model:

\begin{itemize}
  \item The network is trained and tested with the same steganographic algorithm but with different payloads;
  \item The network is trained and tested with the same payload but with different steganographic algorithms.
\end{itemize}

{\setlength{\abovecaptionskip}{2pt}
 \setlength{\belowcaptionskip}{-2pt}
\begin{table}[t]
  \centering
  \renewcommand\arraystretch{1.2}
  \caption{Sensitivity experiment when payloads are mismatched. The S-UNIWARD steganography is used for demonstration. Diagonal elements denote the payload matched case.  }
  \resizebox{9cm}{!} {
  \begin{tabular}{| c | c | c | c | c |  c |}
   \hline
    \diagbox[width=5em,trim=l]{Train}{Test} & 0.05 bpp & 0.1 bpp & 0.2 bpp & 0.3 bpp & 0.4 bpp  \\ \hline
    0.05 bpp & \textbf{42.13\%} & 36.81\% & 31.40\%  & 28.25\% & 26.61\%  \\ \hline
    0.1 bpp & 42.22\% & \textbf{35.21\%} & 28.74\%  & 25.37\% & 23.23\%  \\ \hline
    0.2 bpp & 44.23\% & 36.75\% & \textbf{26.82\%}  & 22.10\% & 19.48\%  \\ \hline
    0.3 bpp & 45.91\% & 39.07\% & 27.85\%  & \textbf{20.71\%} & 16.91\%  \\ \hline
    0.4 bpp & 47.46\% & 41.42\% & 29.97\%  & 21.54\% & \textbf{16.53\%}  \\ \hline
  \end{tabular}
  }
\end{table}}

\begin{figure*}[t]
   \centering
   \begin{subfigure}{.19\textwidth}
     \centering
     \includegraphics[height=3.4cm, width=3.4cm]{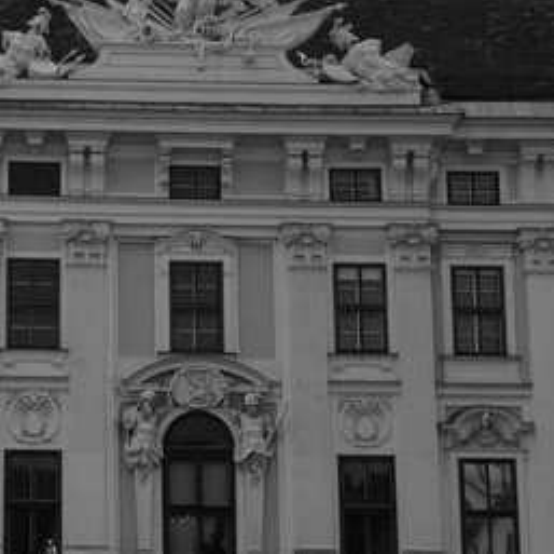}
     \caption{Cover image}
   \end{subfigure}
   \begin{subfigure}{.19\textwidth}
      \centering
      \includegraphics[height=3.4cm, width=3.4cm]{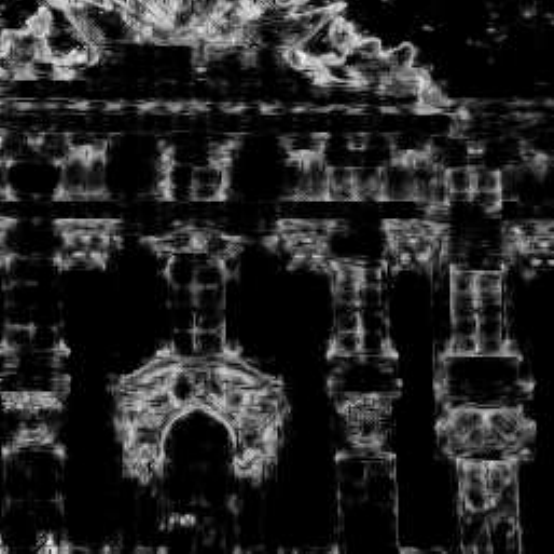}
   \caption{HUGO}
   \end{subfigure}
   \begin{subfigure}{.19\textwidth}
      \centering
      \includegraphics[height=3.4cm, width=3.4cm]{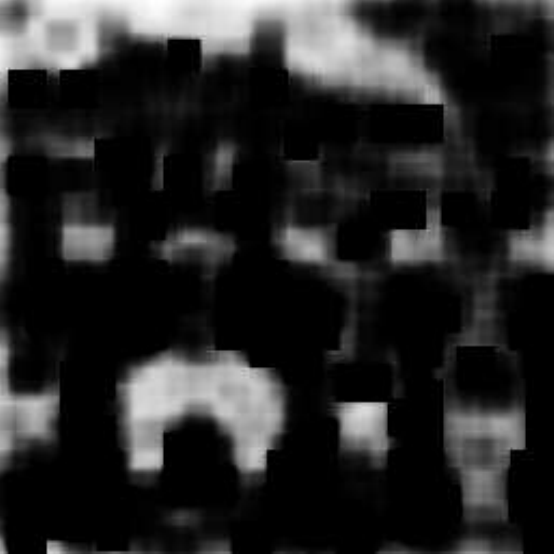}
   \caption{HILL}
   \end{subfigure}
   \begin{subfigure}{.19\textwidth}
      \centering
      \includegraphics[height=3.4cm, width=3.4cm]{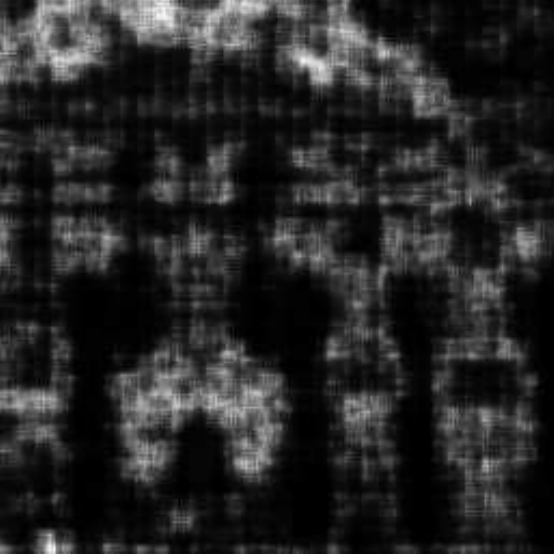}
      \caption{S-UNIWARD}
   \end{subfigure}
   \begin{subfigure}{.19\textwidth}
      \centering
      \includegraphics[height=3.4cm, width=3.4cm]{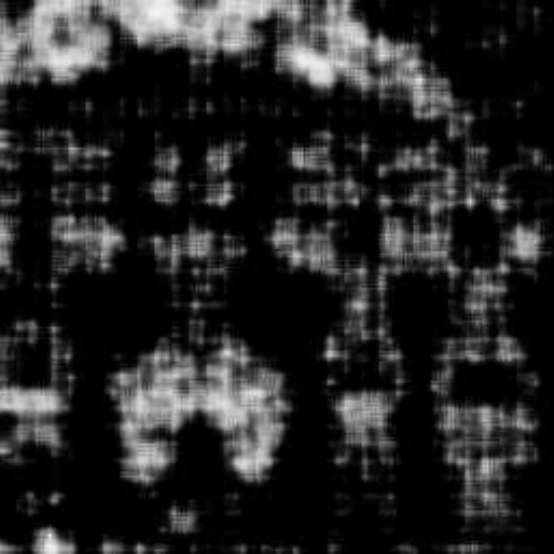}
   \caption{WOW}
   \end{subfigure}
   \caption{A cover image and its embedding probability maps of different steganographic algorithms. (a). a crop of ``13.pgm" cover image from the BOSSbase dataset; (b). the embedding probability of HUGO; (c). the embedding probability of HILL; (d). the embedding probability of S-UNIWARD; (e). the embedding probability of WOW.}
\end{figure*}

\subsubsection{Sensitivity to the Payload Mismatch}
In this experiment, we assess the sensitivity of proposed network by testing it with the payload different from that for network training. Specifically, we first train the proposed network from 0.05 bpp to 0.4 bpp and obtain five optimized networks. Then, each optimized network is validated at these five different payloads. The network for the S-UNIWARD steganography is reported in this subsection. For all cases, 5,000 cover images and their stegos are used for training, while the rest images and their stegos are for testing. In addition, we make sure that cover images in testing are not used to train the network in each payload mismatched case.

Results are reported in the TABLE III. In this table, the payload in each row denotes the payload for network training, while the payload in each column denotes the payload for network testing. The diagonal detection error rates actually denote the case that the training payload and the testing payload are matched. Compared with the payload matched case, we observe that the detection error rate increases gradually as the payload mismatch amplitude increases. More importantly, the loss of detection error rate is quite small when the network is tested at an adjacent mismatched payload, indicating that the proposed network is robust to the payload mismatch.

\begin{figure}[t]
   \centering
   \begin{subfigure}{.45\textwidth}
      \centering
      \includegraphics[height=5cm, width=7cm]{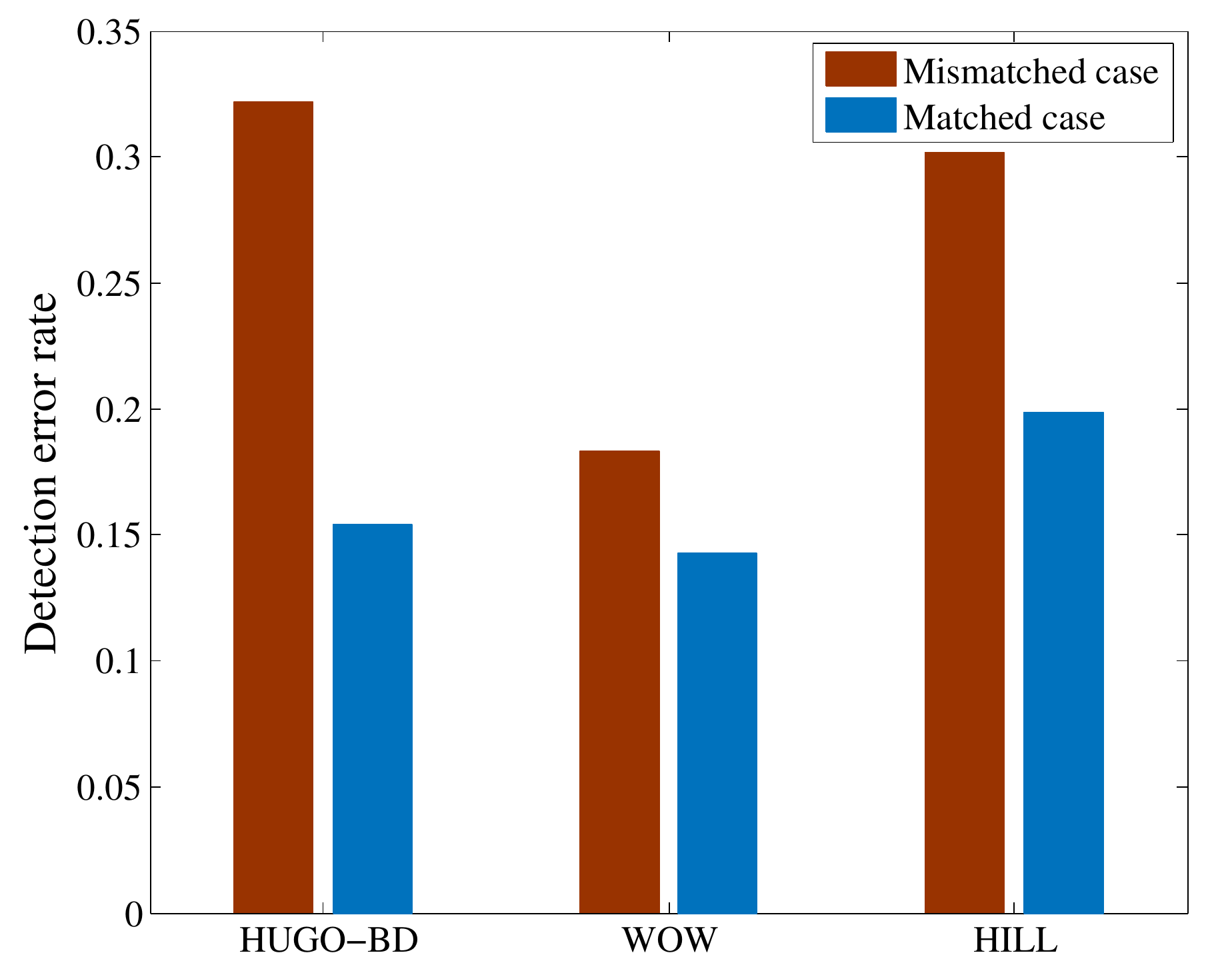}
   \end{subfigure}
   \caption{Sensitivity experiment when steganographic algorithms are mismatched. In mismatched case, the network is trained by S-UNIWARD steganography but tested by other three steganographic algorithms. The training payload and testing payload are fixed at 0.4 bpp.}
\end{figure}

\subsubsection{Sensitivity to the Steganographic Algorithm Mismatch}
In this experiment, we assess the sensitivity of the proposed network by testing it with a steganographic algorithm that is different from the algorithm for the network training. The network, which is trained on the S-UNIWARD steganography, is tested on other three steganographic algorithms. In order to solely investigate whether the proposed network is sensitive to the change of steganographic algorithms, we force that the payload for training and testing is fixed at 0.4 bpp. Same to the setting in the previous experiments, 5,000 cover images and their stegos are used for training, while the rest images and their stegos are used for testing.

Fig.11 shows detection errors of steganographic algorithm matched and mismatched cases. In mismatched cases, we use the network trained on S-UNIWARD steganography to detect other three steganographic algorithms. For HUGO-BD and HILL, the detection error rate of mismatched case degrades significantly to the matched case. The reason is that distortion functions of these two steganographic algorithms for message embedding, which are depicted by Fig.10(a) and Fig.10(b), are different from S-UNIWARD, which is depicted by Fig.10(c). This indicates that secret messages in HUGO-BD and HILL are embedded in completely different patterns to S-UNIWARD, making the network trained by S-UNIWARD hard to detect stego images generated by them. For WOW steganograpy, the performance loss of mismatched case is relatively small because distortion functions used for S-UNIWARD and WOW are similar to each other, as shown in Fig.10(c) and Fig.10(d). Those observations imply that, compared to the payload mismatch, the proposed CNN model is more sensitive to steganographic algorithm mismatch.

\subsection{Data Augmentation and Ensemble Learning}
In the last subsection, we demonstrate whether some techniques widely used in deep learning, including data augmentation and ensemble learning, can improve the detection accuracy of our network. Three settings are conducted here:

\subsubsection{Data augmentation (Aug.)} in this setting, 10,000 BOSSbase samples are randomly split into 5,000 training images and 5,000 testing images. For training images, we rotate them with 90 degree, 180 degree, and 270 degree, generating a new training set with 20,000 samples. Then, the S-UNIWARD steganography is used to embed secret messages into the augmented training set and the test set. The proposed network is trained on this new training set with 20,000 covers/stegos and finally validated on the test set with 5,000 covers/stegos.

\subsubsection{Ensemble learning (Ens.)} similar to the first setting, we first split the BOSSbase dataset into 5,000 training images and 5,000 test images randomly. Then, we use the S-UNIWARD steganography to embed secret messages into images and obtain 5,000 training pairs and 5,000 testing pairs. Then, we generate five new training sets by randomly selecting image pairs from the original training sets, where each new set contains 4,000 pairs. Finally, five CNN models are trained based on new training sets, and they are tested on the original test pairs. The overall performance of ensemble learning is obtained by voting five optimized models.

\subsubsection{Data augmentation + ensemble learning (Aug. + Ens.)} same to the first setting, a training set with 20,000 pairs and a test set with 5,000 pairs are created. Same to the second setting, five new training sets are generated by random selection, where each set contains 16,000 pairs. Then, five CNN models are trained based on five augmented training sets respectively. The overall detection accuracy is obtained by voting these optimized CNN models.

{\setlength{\abovecaptionskip}{2pt}
 \setlength{\belowcaptionskip}{-2pt}
\begin{table}[t]
  \centering
  \renewcommand\arraystretch{1.2}
  \caption{Detection error rates of the proposed network with data augmentation and ensemble learning. The S-UNIWARD steganography at payload 0.4 bpp is used for validation. }
  \resizebox{8.7cm}{!} {
  \begin{tabular}{| c | c | c | c | c |}
   \hline
  \textbf{Method} & Original & Aug. & Ens. & Aug. + Ens. \\ \hline
  Single (average)& 16.53\% & 14.86\% & 18.23\% & 15.36\% \\ \hline
  Ensembled & $-$ & $-$ & 16.79\% & \textbf{14.07\%} \\ \hline
  \end{tabular}
  }
\end{table}}

The detection error rates of three schemes are reported in TABLE IV. Compared with the original model, data augmentation, ensemble learning and their combination can improve the performance of the proposed network. These results indicate that those deep learning techniques are indeed beneficial for CNN based steganalysis.

\section{Conclusion}
Even though convolutional neural network based steganalysis develops a lot in recent years, what kinds of deep learning techniques suitable for image steganalysis wait to be carefully examined. In this paper, we deeply explore the batch normalization, a popular technique used in general image classification, for the steganalysis task. Theoretically, we analyze that a CNN model with multiple batch normalization layers can detect steganography at very high accuracy when cover images and their stegos are paired training and testing, and their batch statistics are obtained to normalize the data. However, the network becomes unstable and fails to detect stego images when cover-setgos are unpaired. To handle this difficulty, we propose a novel normalization technique called shared normalization to normalize input data. Compared with the batch normalization, the proposed normalization method can stabilize network learning and make the network obtain better generalization ability. Based on the SN, we further propose a novel CNN model for image steganalysis. The experiment demonstrates that the proposed model achieved better performances than traditional rich model method and state-of-the-art CNN models. In future works, we would extend our model to detect steganography in compressed domain.

\appendices
\section{Paired Training but Testing with Fixed Statistics}
From Eq.(17) and Eq.(18), we can conclude that the model learning is actually to maximize the following difference:
\begin{equation}
  d_{n} = \left| E\left[ \mathbf{f}^{ps}_{n} \right] -  E\left[ \mathbf{f}^{pc}_{n} \right] \right| = \left| \frac{E\left[ \mathbf{W}_{n}\mathbf{s}_{n-1}^{op} \right]}{\sigma_{n}} \right|
\end{equation}
to maximize Eq.(36), the network should update $\mathbf{W}_{n}$ to the direction that increase $d_{n}$. Here, $d_{n}$ denotes the margin between the feature map of cover image $E\left[ \mathbf{f}^{ps}_{n} \right]$ and the feature map of stego image $E\left[ \mathbf{f}^{pc}_{n} \right]$. In addition, whether the network with multiple batch normalization layers can detect steganography well depends on $d_{n}$. If the data is normalized with accurate statistics, the model can classify covers and stegos accurately. Otherwise, the network fails to detect steganography when the statistics is not accurate.

The performance of a network with multiple batch normalization layers would vibrate if estimated batch statistics $\mu$ and $\sigma$ rather than the accurate batch statistics are used in testing phase. In the fixed case, the output of a cover image $\mathbf{x}$ after two ``Conv+BN+ReLU" blocks is:
\begin{equation}
 \begin{aligned}
  \mathbf{x}_{2}^{o} & = \frac{\mathbf{W}_{2}(\mathbf{W}_{1}\mathbf{x} - \mu_{1})}{\sigma_{1}\sigma_{2}} \circ \mathcal{H}\left(\frac{\mathbf{f}^{o}_{1}}{\sigma_{1}}\right) \circ \mathcal{H}\left(\frac{\mathbf{f}^{o}_{2}}{\sigma_{2}}\right) \\
   & - \frac{\mu_{2}}{\sigma_{2}} \circ \mathcal{H}\left(\frac{\mathbf{f}^{o}_{2}}{\sigma_{2}}\right)
 \end{aligned}
\end{equation}
where $\mu_{1}, \mu_{2}, \sigma_{1}, \sigma_{2}$ are fixed parameters of the batch normalization layer, $\mathbf{f}^{o}_{1}$ and  $\mathbf{f}^{o}_{2}$ are defined as:
\begin{equation}
  \mathbf{f}^{o}_{1} = \mathbf{W}_{1}\mathbf{x} - \mu_{1}
\end{equation}
\begin{equation}
  \mathbf{f}^{o}_{2} = \mathbf{W}_{2}\mathbf{x}_{1}^{o} - \mu_{2}
\end{equation}

According to the analysis, the discrimination of a cover image and a stego image depends on the difference $d_{n}$. The estimated normalization statistics fall into the region (we use the mean as an example):
\begin{equation}
     |\mu_{i}-\mu_{\mathcal{B}}| > d_{n} \ \ \ i=1,2
\end{equation}
where $\mu_{\mathcal{B}}$ denotes the accurate mean of a mini-batch. In this case, the bias introduced by inaccurate $|\mu_{i}-\mu_{\mathcal{B}}|$ is larger than the margin $d_{n}$, which will be accumulated in deep layers and finally makes the network fail to detect steganography well. $d_{n}$ is a small value since elements in $\mathbf{s}_{n-1}^{op}$ are small. Therefore, Eq.(40) can be satisfied easily if the estimated statistics are not same to the batch statistics.

\ifCLASSOPTIONcaptionsoff
  \newpage
\fi

%
%


\end{document}